   \documentclass[aps,twocolumn,prb,showpacs]{revtex4}

\usepackage{graphicx}
\usepackage{latexsym}
\usepackage{calc}
\usepackage{amsmath}

\begin{document}

\title{Ordered phases and phase transitions in \\ the fully
frustrated XY model on a honeycomb lattice}

\author{S.~E. Korshunov}
\affiliation{L.~D.~Landau Institute for Theoretical Physics RAS, 142432
Chernogolovka, Russia}

\date{December, 2011}


\begin{abstract}
The phase diagram of the fully frustrated XY model on a honeycomb lattice
is shown to incorporate three different ordered phases. In the most
unusual of them, a long-range order is related not to the dominance of a
particular periodic vortex pattern but to the orientation of zero-energy
domain walls separating domains with different orientations of vortex
stripes. The phase transition leading to the destruction of this phase can
be associated with the appearance of free fractional vortices and is of
the first order. The stabilization of the two other ordered phases
(existing at lower temperatures) relies a positive contribution to
domain-wall free energy induced by the presence of spin waves. This effect
has a substantial numerical smallness, in accordance with which these two
phases can be observed only in the systems of really macroscopic sizes.
In physical systems (like magnetically frustrated Josephson junction
arrays and superconducting wire networks), the presence of additional
interactions must lead to a better stabilization of the phase with the
long-range order in terms of vortex pattern and improve the possibilities
of its observation.

\end{abstract}

\maketitle

\section{Introduction}

A uniformly frustrated XY model can be defined by the Hamiltonian
\begin{subequations}                                         \label{FFXY}
\begin{equation}
H=\sum_{(jj')}V(\varphi_{j'}-\varphi_{j}-A_{jj'})\;,          \label{H}
\end{equation}
where the interaction of the variables $\varphi_{j}$  defined on the sites
$j$ of some regular two-dimensional lattice is described by periodic
function
\begin{equation}                                              \label{V}
    V(\theta)=-J\cos\theta\;,
\end{equation}
$J>0$ is the coupling constant,  and the summation is performed over all
pairs of nearest neighbors $(jj')$ on the lattice. The non-fluctuating
(quenched) variables $A_{jj'}\equiv -A_{j'j}$ defined on lattice bonds
have to satisfy the constraint
\begin{equation}                                            \label{f}
\sum_{\mbox{\small\raisebox{-0.6mm}{$\Box\,$}}\alpha} A_{jj'}=2\pi
f\,(\mbox{mod}\,2\pi)
\end{equation}
\end{subequations}
on all  lattice plaquettes. The notation
$\mbox{\raisebox{-0.4mm}{$\Box$}}\hspace*{0.7mm}\alpha$ below the sign of
summation implies the directed sum of variables $A_{jj'}\equiv -A_{j'j}$
over the perimeter of plaquette $\alpha$ in the counterclockwise
direction. For $f=0$ the model defined by Eqs. (\ref{FFXY}) is isomorphic
to the conventional XY model (without frustration), whereas for
$f=\frac{1}{2}$ (the maximal irreducible value of $f$, see Ref.
\onlinecite{TJ-L}) this model is called fully frustrated.

Since eighties, the uniformly frustrated and, especially, the fully
frustrated XY models on various lattices have been studied rather
intensively (for reviews, see Refs. \onlinecite{HPV} and
\onlinecite{ufn}), mostly in relation with experiments on Josephson
junction arrays. \cite{rev} In these artificial superconducting systems
variables $\varphi_{j}$ can be associated with the phases of the
superconducting order parameter on different superconducting grains
forming an array, and $A_{ij}$ is related to the vector potential of a
uniform magnetic field, whose
magnitude corresponds to having $f$ 
superconducting flux quanta per lattice plaquette. The form of Eq.
(\ref{f}) corresponds to taking into account only the external magnetic
field and neglecting the field of weak currents flowing in the junctions.
Planar magnets with odd number of antiferromagnetic bonds per plaquette
\cite{Vil} are also described by the fully frustrated XY models. The
recent renewal of the interest to the fully frustrated Josephson junction
arrays has been related to their possible application for the creation of
topologically protected quantum bits. \cite{IF}

In a typical situation, the energy of each of the Josephson junctions
forming an array as a function of the gauge-invariant phase difference
\begin{equation}
\theta_{jj'}=\varphi_{j'}-\varphi_j-A_{jj'}
\end{equation}
is indeed rather accurately described by the function (\ref{V}). On the
other hand, a magnetically frustrated network of thin superconducting
wires can be described \cite{ufn} in the London limit (when the amplitude
of the superconducting order parameter can be assumed to be constant along
the wires) by the same Hamiltonian (\ref{H}) with $V(\theta)$ replaced by
the so-called Berezinskii-Villain interaction \cite{Ber,Vil75}
having the same symmetry and periodicity as $V(\theta)$, see Appendix
\ref{BV}. The fully frustrated XY model with the Berezinskii-Villain
interaction 
is more convenient for a theoretical analysis, because its partition
function can be rigorously transformed into that of a Coulomb gas with
half-integer charges \cite{FHS} (Appendix \ref{BV} gives more details). In
a number of cases it can be expected that the main features of the
frustrated XY models with the conventional and with the
Berezinskii-Villain interaction are the same.

The ground states of the fully frustrated XY models are characterized by
the coexistence of the continuous $U(1)$ degeneracy (related to the
possibility of the simultaneous rotation of all phases) with a discrete
one, whose nature depends on the structure of the lattice.
Accordingly, the fully frustrated XY models allow in addition to the phase
transition related to unbinding of vortex pairs also for the existence of
other phase transitions related to the disordering of the discrete degrees
of freedom. The main question in such a situation is what is the sequence
and the nature of these phase transitions.

The most thoroughly studied examples of the fully frustrated XY models are
the models on square and triangular lattices. In both of them the ground
states are characterized by a regular alternation of the plaquettes with
positive and negative vorticities. Accordingly, the discrete degeneracy of
the ground states is two-fold, that is, the simplest one which is
possible. However, it took about two decades before it was firmly
understood \cite{ufn,K02} why the Berezinskii-Kosterlitz-Thouless
transition related to unbinding of vortex pairs has to take place in these
models at lower temperatures than the second phase transition related to
the proliferation of the Ising-type domain walls. Due to a strong mutual
influence of the two transitions (which are situated rather close to each
other), a convincing numerical demonstration of the Ising nature of one of
them has required very substantial efforts. \cite{HPV,Ols97} Before that
quite a number of works appeared claiming that this transition has a
non-Ising critical behavior, which later has been demonstrated to be a
finite-size effect.


The properties of the fully frustrated XY model on a a honeycomb lattice
are not so well understood. The family of its ground state is known to
have an infinite discrete degeneracy, \cite{ShS85} which can be
conveniently described in terms of the formation of zero-energy domain
walls parallel to each other. \cite{K86,KD} The analysis of the
half-integer Coulomb gas on a triangular lattice (the dual representation
of the considered XY model) has revealed \cite{LT} that in this system the
formation of a domain with a different charge pattern can cost only a
finite energy even when the size of this domain is arbitrary large. This
has led the authors of Ref. \onlinecite{LT} to the conclusion on the
absence of a long-range order at any nonzero temperature $T$.

This conclusion is not directly applicable to the fully frustrated XY
model with interaction (\ref{V}) 
because at $T>0$ the zero-energy domain walls acquire a positive free
energy $f_{\rm DW}$ originating from the difference in the free energy of
the continuous fluctuations (spin waves). \cite{KD} However, the effect is
very weak and according to the estimates in Ref. \onlinecite{KD} can
manifest itself only if the size of the system $L$ substantially exceeds
$L_c\gtrsim 10^5$, which makes it unobservable in real and numerical
experiments on systems with $L\lesssim L_c$.

Nonetheless, up to now it has remained unclarified if in situations when
$f_{\rm DW}$ is absent (as it happens in the case of the
Berezinskii-Villain interaction) or can be neglected, the system can still
demonstrate a phase transition related 
to the continuous degrees of freedom or the destruction of the long-range
order of in terms of vortex pattern prevents the possibility of such a
transition. The available numerical data \cite{ShS84,ShS85,LT,RBM} does
not lead to a unique conclusion and allows for different interpretations,
including the existence of a spin-glass transition. \cite{RBM}

In this work we reexamine the fully frustrated XY model on a honeycomb
lattice with the aim of finding an answer to this question, as well as
establishing what is the nature of the phase transition (transitions)
induced by the positiveness of $f_{\rm DW}$. Our main conclusions can be
formulated as follows.

Even when $f_{\rm DW}$ is equal to zero (or can be neglected), this does
not mean the complete absence of the long-range order at any nonzero
temperature. In a such a situation, the long-range order is related not to
the dominance of a particular vortex pattern, but to a preferable
orientation of the zero-energy domain walls.

The phase transition leading to the destruction of this long-range order
is related to the appearance of a finite concentration of unpaired
fractional vortices with topological charges $q=\pm\frac{1}{8}$. The
fractional vortices are located at the nodes of the domain-wall network
existing in the system with $f_{\rm DW}=0$ at any temperature, however at
low temperatures they have to be bound in pairs, which imposes the
existence of a preferable orientations of domain walls.

In the model with the conventional interaction
\makebox{$V(\theta)=-J\cos\theta$} which at $T>0$ has small but positive
$f_{\rm DW}$ (induced by spin-waves), the increase of temperature leads to
the sequence of three phase transitions. However, the two of them require
for their observation very large systems with $L\gg L_c\sim 10^7$. In
physical systems like magnetically frustrated Josephson junction arrays
and superconducting wire networks, the value of $L_c$ will be lower due to
the presence of other mechanisms for the removal of the accidental
degeneracy.

Although the lowest-temperature phase transition (whose critical
temperature tends to zero  when $f_{\rm DW}\rightarrow 0$) consists in the
appearance of domain walls, it does not lead to the destruction of the
long-range order in terms of vortex pattern. On the contrary, it is
related to a partial restoration of the continuous $U(1)$ symmetry. This
phase transition belongs to the Ising universality class, whereas the two
other transitions mentioned above must be of the first order.

A more detailed summary of the phase diagram of the fully frustrated XY
model on a honeycomb lattice and of the properties of different phases and
phase transitions is presented in the concluding Sec. \ref{conc}. It also
compares our results with that of other works and discusses their
relevance for some experimental situations and other uniformly frustrated
XY models that allow for the formation of parallel zero-energy domain
walls.

\section{Zero temperature analysis}
\subsection{The ground states \label{sec:gs}}

When speaking about both global and local minima of the Hamiltonian of an
XY model it is convenient to characterize them in terms of vorticities of
different plaquettes. Vorticity $v_\alpha$ of plaquette $\alpha$ can be
defined as the directed sum  over the perimeter of this plaquette of the
gauge-invariant phase differences
\makebox{$\theta_{jj'}=\varphi_{j'}-\varphi_j-A_{jj'}$} 
reduced to the interval $(-\pi,\pi)$. Here and below we use Greek indices
for labelling the plaquettes, each of which can be associated with a
particular site of the triangular lattice ${\cal T}$ dual to the original
honeycomb lattice $\cal H$.

In the fully frustrated XY model on a honeycomb lattice variables
$v_\alpha$ can acquires values $\pm \pi,\pm 3\pi,\pm 5\pi$. Since the main
role in the thermodynamics is played by the minima of Hamiltonian
(\ref{FFXY}) with $v_a=\pm \pi$ (as well as by fluctuations in the
vicinities of these minima), one often speaks not about vorticities but
about chiralities \makebox{$\sigma_\alpha=v_\alpha/\pi=\pm 1$} of
different plaquettes. The knowledge of the chiralities of all plaquettes
is sufficient for restoring (up to a global rotation) the values of phase
variables $\varphi_j$ in the minimum of the Hamiltonian which can be
associated with this particular configuration of chiralities.

In all ground states of the fully frustrated XY model on a honeycomb
lattice the plaquettes with different signs of chirality are separated by
the bonds with
\makebox{$\theta_{jj'}
=\pm\pi/4$}, whereas the plaquettes with the same sign of chirality are
separated by the bonds with $\theta_{jj'}=0$. In Fig. \ref{GroundStates}
showing some of the ground states the plaquettes with positive (negative)
chirality are shaded (unshaded), whereas the bonds with
$\theta_{jj'}=\pm\pi/4$ are shown by arrows. On any bond with an arrow,
$\theta_{jj'}$ is equal to $+\pi/4$ if the arrow is directed from site $j$
to site $j'$ and to $-\pi/4$ in the opposite case.

Fig. \ref{GroundStates}(a) shows an example of the ground state with the
simplest possible structure. In this state the plaquettes with the same
sign of vorticity form parallel straight stripes. Accordingly, such states
are called below the striped states. The set of striped states is
characterized by a sixfold discrete degeneracy related to the
possibilities of shifting and rotating vortex pattern. Naturally, it also
has the continuous $U(1)$ degeneracy related to the simultaneous rotation
of all phases, but in the following the term ``striped state'' will always
refer just to the configuration of chiralities, without specifying the
phases.

Striped states allow for the formation of domain walls (separating
different realizations of such states) which cost no energy. \cite{K86} An
example of such a zero-energy domain wall (ZEDW) is shown in Fig.
\ref{GroundStates}(b). An arbitrary number of ZEDWs separated by arbitrary
distances can be introduced into the system in parallel to each other, as
is shown in Fig. \ref{GroundStates}(c).

The only possible intersection of ZEDWs which does not cost extra energy
is shown in Fig. \ref{GroundStates}(d). Since each ZEDW can be ascribed a
particular direction in accordance with the orientation of the angle made
by stripes, as shown in Fig. \ref{GroundStates}, it is clear that in a
ground state the system cannot contain more than one intersection of such
a kind and in the case of the periodic boundary conditions even one is
impossible. Therefore, a typical ground state contains an irregular
sequence of straight ZEDWs which are parallel to each other, as shown in
Fig. \ref{GroundStates}(c). The presence of a boundary between two ground
states with a finite concentration of ZEDWs and different orientations of
these walls would cost an energy proportional to the length of this
boundary.

The structure of the ground states 
is not sensitive to a particular form of the interaction in the
Hamiltonian of the model, as soon as it is even in $\theta$ and behaves
more or less in the same way as $V(\theta)=-J\cos\theta$ [that is, has the
single maximum at $\theta=\pi$ and the single minimum at $\theta=0$ whose
width is comparable with the period of $V(\theta)$]. For example, it can
be also the Berezinskii-Villain interaction defined by Eq. (\ref{V_BV}) or
the piecewise parabolic interaction (the zero-temperature limit of the
Berezinskii-Villain interaction).
\begin{widetext}

\begin{figure}[h]
\vfill
\includegraphics[width=160mm]{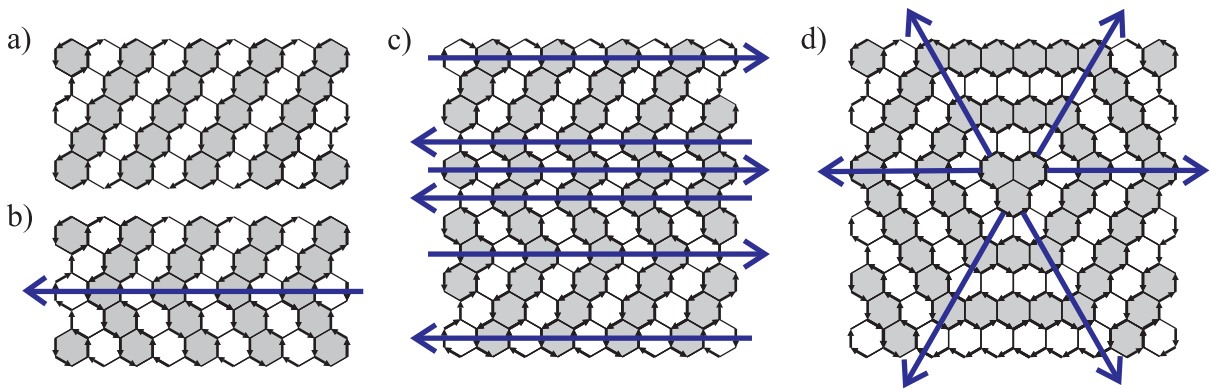} \caption{(Color online)
The ground states of the FFXY model on a honeycomb lattice; (a) a
striped state; 
(b) a state with a single zero-energy domain wall separating two different
striped states; (c) a typical ground state with an irregular sequence of
parallel domain walls; (d) the only possible intersection of domain walls
which does not cost extra energy. The plaquettes with positive (negative)
chirality are shaded (unshaded).} \label{GroundStates}
\end{figure}

\newpage
\end{widetext}

\subsection{Zero-energy domain walls and long-range order\label{T=0-LRO}}

If domain walls separating different striped states would have a positive
energy per unit length, at zero temperature the system would be frozen in
one of the six striped states. In terms of chiralities $\sigma_\alpha$,
the order parameter corresponding to the corresponding long-range order
(LRO) can be written as
\begin{equation}                                                \label{M}
    S=\sum_{n=1}^{3}e^{i\frac{2\pi n}{3}}
    \sum_{\alpha}s_{n,\alpha}\,,~~~
    s_{n,\alpha}=\sigma_\alpha \exp
    \frac{{i\bf Q}_n{\bf r}_\alpha}{2}\,.
\end{equation}
Here and below ${\bf r}_\alpha$ denotes the positions of the sites
$\alpha$ of the dual triangular lattice ${\cal T}$ which can be associated
with the plaquettes of the original honeycomb lattice $\cal H$, whereas
${\bf Q}_n$ (with $b=1,2,3$ and ${\bf Q}_1+{\bf Q}_2+{\bf Q}_3=0$) are the
three reciprocal vectors of $\cal T$. In any of the striped states one of
the three sums $S_n\equiv\sum_{\alpha}s_{n,\alpha}$ grows with size of the
system as $S_n=\pm N$ ($N$ being the total number of the plaquettes in the
system), whereas the two other are equal to zero. The correlation function
\begin{equation}                                                 \label{Cch}
    C_{\rm ch}({\bf r}_\alpha-{\bf r}_\beta)=\langle\sigma_\alpha
    \sigma_\beta\rangle\;,
\end{equation}
where the average is taken over the six striped vortex patterns, has an
oscillating behavior. Namely, if triangular lattice $\cal T$ is
partitioned into four equivalent triangular sublattices, then $C_{\rm
ch}({\bf r}_\alpha-{\bf r_\beta})$ is equal to 1 when $\alpha$ and $\beta$
belong to the same sublattice and to $-1/3$ when they belong to different
sublattices.

When the domain walls shown in Fig. \ref{GroundStates} cost no energy, the
oscillating behavior (without a decay at \makebox{$|{\bf r}_\alpha-{\bf
r}_\beta|\to\infty$}) of the correlation function (\ref{Cch}) (where the
average now has to be taken over the infinite set of the ground states) is
retained only when $\alpha$ and $\beta$ belong to the same column of sites
on $\cal T$, that is, when vector ${\bf r}_\alpha-{\bf r}_\beta$ is
directed along ${\bf e}_n$, one of the three lattice vectors of $\cal T$
(${\bf e}_1+{\bf e}_2+{\bf e}_3\equiv 0$). In such a case, in one third of
the ground states the line connecting $\alpha$ and $\beta$ cannot be
crossed by any domain walls and therefore the product
$\sigma_\alpha\sigma_\beta$ depends only on the distance between $\alpha$
and $\beta$. For all other directions of ${\bf r}_\alpha-{\bf r}_\beta$,
the correlations of $\sigma_\alpha$ and $\sigma_\beta$ are absent, which
leads to
\begin{equation}                                               \label{S^2}
\langle |S|^2\rangle\propto N\;.
\end{equation}
This means that the long-range order in terms of $s_{n,\alpha}$ is
destroyed. It is worthwhile to emphasize explicitly that the dependence
(\ref{S^2}) is not a consequence of algebraic correlations of
$\sigma_\alpha$ as it could seem from its form, but has a different origin
explained above.

However, it is not hard to note that the destruction of the long-range
order is not complete. In the presence of a random sequence of straight
parallel ZEDWs, the system contains the domains of only four different
striped states out of six, whereas the two other vortex patterns (with
stripes parallel to the direction of the domain walls) are not represented
at all. Therefore, in a typical ground state the system still possesses a
long-range order. The most evident interpretation of this order consists
in associating it with the direction of ZEDWs.

Instead of really monitoring the direction of domain walls (whose
identification requires analyzing chirality patterns in rhombic clusters
formed by four neighboring plaquettes) an order parameter sensitive to the
direction of domain walls can be introduced
by studying the distribution of energy between the bonds with different
orientations. In a typical ground state the average energy of the bonds
whose direction is perpendicular to the orientation of domain walls is
equal to $V(\pi/4)$, whereas for the bonds with two other orientations it
is equal to $\frac{1}{2}[V(0)+V(\pi/4)] < V(\pi/4)$. This allows one to
introduce an order parameter as
\begin{equation}                                                
    D=\sum_{(jj')}\exp\left[i\frac{2\pi n_{jj'}}{3}\right]V(\theta_{jj'})\;,
\end{equation}
where the value of $n_{jj'}=1,2,3$ depends on the orientation of the bond
$(jj')$.

Thus, the vanishing of the energy of domain walls does not lead to the
complete destruction of the long-range order in the system and one can
expect that there should occur a phase transition related to the
destruction of the long-range order described by order parameter $D$. Note
that $D$  is proportional to the area of the system also when one of the
six striped vortex pattern is a dominant one. However, in contrast to $S$,
order parameter $D$ does not allow to distinguish between two vortex
patterns which differ from each other by changing the signs of all
chiralities.

\vspace*{-3mm}
\subsection{Phase correlations}
\vspace*{-1mm}

When speaking about phase correlations, one should not forget that a
change of gauge leads to the multiplication of $\exp
[i(\varphi_\alpha-\varphi_\beta)]$ by some phase factor. In accordance
with that, a gauge-invariant description of phase correlations can be
achieved by paying attention only to the absolute values of the standard
correlation functions. For taking into account the peculiarities of the
considered model, it is convenient to introduce the set of gauge-invariant
phase correlation functions defined as
\begin{equation}                                              \label{C_p}
C_p({\bf r}_j-{\bf r}_k)=
    |\langle\exp [ip(\varphi_j-\varphi_k)]\rangle|\,
\end{equation}
and numbered by positive integer $p$. Another approach to introducing
gauge-invariant phase correlation functions involves considering two
identical but completely independent replicas of the system. \cite{BY}

If domain walls separating different striped states would have a positive
energy per unit length, at zero temperature there would exist a true
long-range order in terms of phase variables. In particular, if the
honeycomb lattice is partitioned into $32$ equivalent triangular
sublattices, for any two sites $j$ and $k$ belonging to the same
sublattice, $C_p({\bf r}_j-{\bf r}_k)$ would be equal to $1$ for any
integer $p$. On the other hand, when $p$ is a multiple of eight ($p=8p'$
with integer $p'$), $C_{p}({\bf r}_j-{\bf r}_k)$  would be equal to $1$
for any pair of sites. This is evident already from the fact that for
$\theta_{jj'}=0,\pm \pi/4$ the factors $\exp i(8\theta_{jj'})$ are always
equal to $1$.

For evident reasons, the presence of a random sequence of straight domain
walls separating different striped states  cannot change the behavior of
$C_{p}({\bf r})$ with $p=8p'$, which remain identically equal to $1$.
However, it leads to an exponential decay of $C_p({\bf r})$ for $p\neq
8p'$ for all directions except those  parallel to ${\bf e}_k$. For these
special directions, all phase correlation functions with $p<8$ have an
oscillating behavior. Like with chiralities, when one calculates the
averages of the squares of certain Fourier transforms of $\exp
(ip\varphi_j)$ over all grounds states, they diverge in a typical ground
state proportionally to the area of the system, but this should not be
taken for the signature of the algebraic correlations of $\exp
(ip\varphi_j)$.


\section{Removal of the accidental degeneracy by spin waves
\label{SpinWaves}}

The infinite degeneracy of the ground states which manifests itself
through the possibility of the formation of ZEDWs is of the accidental
nature ({\em i.e.}, it is not imposed by the symmetries of the system).
Accordingly, it can removed by the addition of some small interactions
preserving the symmetries of the Hamiltonian. It is also removed at a
finite temperature when one takes into account the free energy of the
continuous fluctuations (spin waves). This mechanism for the removal of an
accidental degeneracy \cite{VBCC,Shen} is often referred to as ``order
from disorder". In systems with a continuous degeneracy it is usually
sufficient to compare the contributions from the harmonic fluctuations.
\cite{Shen,Kaw,KVB}

In contrast to that, in the fully frustrated XY model on a honeycomb
lattice the free energy of the harmonic fluctuations is exactly the same
for all ground states of the model. \cite{KD} This is a consequence of a
hidden symmetry existing in the Hamiltonian describing such fluctuations.
\cite{FF-dice} The difference in the spin-waves free energy appears only
when one goes beyond the harmonic approximation. The analysis of the
third- and the fourth-order anharmonic terms reveals that the spin-wave
free energy is minimal for the striped states whereas the presence of
ZEDWs adds to the free energy of the system a positive term roughly
proportional to the total length of the walls.  \cite{KD}  This allows one
to ascribe to domain walls the free energy per unit length
\begin{equation}                                              \label{fDW}
f_{\rm DW}=\gamma T^2/J,~~~~~\gamma\approx 0.7\times 10^{-4}\,,
\end{equation}
where by unit length we mean the lattice constant of the dual triangular
lattice  $a_\triangle$. In terms of $a_\triangle$, the lengths of all
walls are integer.

The positiveness of $f_{\rm DW}$ suggests that in the thermodynamic limit
the free energy of a domain wall crossing the whole system is infinite and
therefore at the lowest temperatures (at which the positiveness of $f_{\rm
DW}$ is not killed by other types of fluctuations)
such walls must be absent. This ensures the existence of the long-range
order corresponding to the dominance of one of the six equivalent striped
vortex patterns. On the other hand, at any nonzero temperature the
presence of spin waves leads to an algebraic decay of the correlation
functions describing phase correlations.

In terms of chiralities $\sigma_\alpha=\pm 1$ (defined on sites $\alpha$
of the dual triangular lattice) the structure of the striped state is
exactly the same as that of the ground state of the antiferromagnetic
Ising model on a triangular lattice with an additional weak
antiferromagnetic interaction of second neighbors. \cite{Metc} In this
model there exist two different scenarios for the disordering the ordered
state, \cite{K05} the choice between which depends on the relations
between the coupling constants. The basic scenario consist in having a
single first-order transition leading to the formation of an isotropic
domain wall network mixing the domains of all six striped states. This
leads to the complete loss of the long-range order. It is interesting that
this phase transition takes place when the free energy of a domain wall
(per unit length) is still positive and not when it vanishes.

However, for some relations between the coupling constants the first-order
transition related to the formation of a domain wall network can be
preceded by a continuous phase transition related to vanishing of the free
energy of double domain walls. This phase transition leads only to a
partial loss of the long-range order. Namely, above it the threefold
degeneracy with respect to the directions of stripes still persists, but
the system consists of alternating domains of the two striped states with
the same orientation of the stripes. \cite{K05}

The existence of different options suggests that in the fully frustrated
XY model on a honeycomb lattice, the conclusion on how the disordering of
the striped state takes place also should be based on the comparison of
different mechanisms for disordering.  In particular, it still remains to
be established what is the nature of the first (with the increase in
temperature) phase transition induced by the extreme smallness of the
domain wall free energy and whether it leads to the complete loss of the
long-range order, or only to its partial loss which leaves a place for
another phase transition (or transitions).

In order to answer these questions we have to understand what is the most
efficient mechanism for the proliferation of domain walls. And since the
decrease of the free energy of domain walls is induced by the presence on
them of point-like defects we have to find the point-like defects which
play the dominant role in the development of the domain-wall fluctuations.

\section{POINT-LIKE DEFECTS \label{PointLikeDefects}}

\subsection{Conventional vortices and vortex pairs}

As any other XY model, the fully frustrated XY model on a honeycomb
lattice allows for the formation of conventional vortices with integer
topological charges. In particular, a vortex with topological
charge $q=\pm 1$ is created when the sign of chirality of some plaquette
is flipped in comparison with what it would be in a ground state. On any
contour surrounding the core of a vortex with topological charge $q$ the
deviations of gauge-invariant phase differences
$\theta_{jj'}=\varphi_{j'}-\varphi_j-A_{jj'}$ from the values they would
have in a ground state ($0$ or $\pm \pi/4$) sum up to $2\pi q$.

At low temperatures vortices with integer topological charges can be
present in the system only in the form of small neutral pairs whose
concentration at $T\ll J$ has to be exponentially low. Below we
demonstrate that all phase transitions in the considered model take place
at $T\ll J$, when the bound pairs of integer vortices are exponentially
rare and, accordingly, have no influence on the properties of the system.
For this reason the presence of such pairs can be neglected.

In addition to conventional vortices, in frustrated XY models there also
can exist fractional vortices localized on corners and intersections of
domain walls. \cite{ufn} In situation when the energy of domain walls
connecting different fractional vortices is exactly equal to zero, the
energies of neutral pairs of fractional vortices will be much lower than
that of pairs of conventional vortices. Accordingly, one can then expect
the neutral pairs of fractional vortices to be the most important
finite-energy excitations at the lowest temperatures.

\subsection{Fractional vortices \label{FV}}

\begin{widetext}

\begin{figure}[h]
\includegraphics[width=130mm]{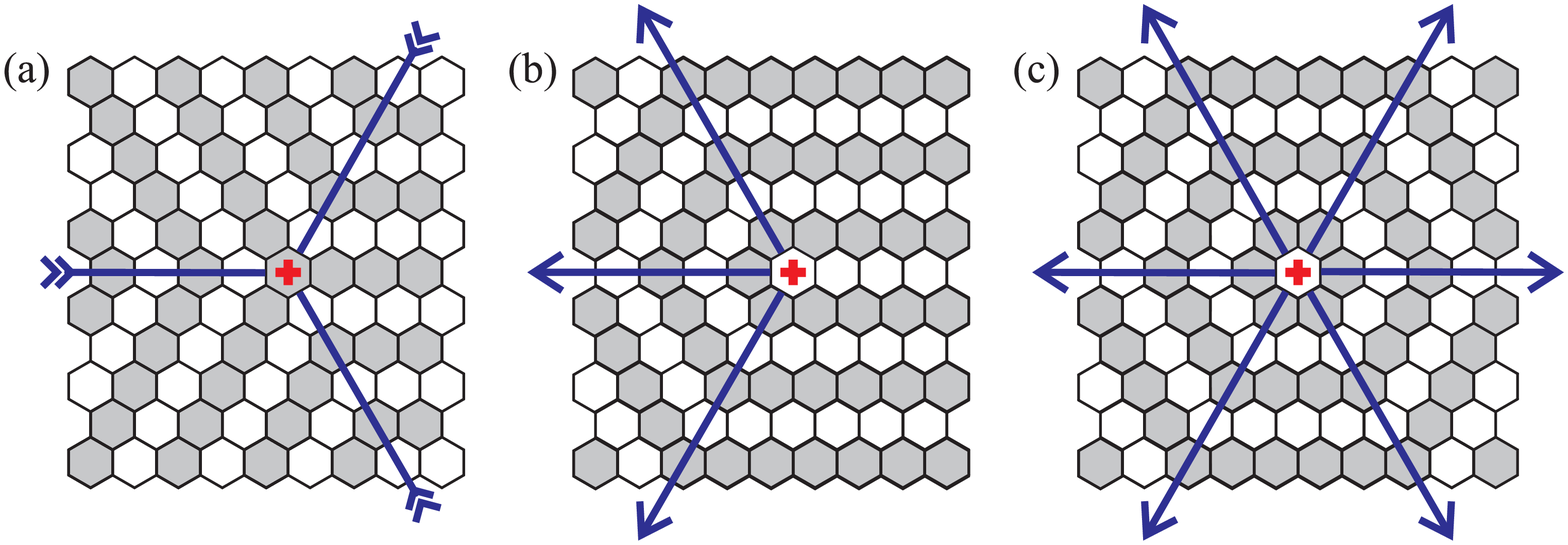} \caption{(Color online)
The simplest point-like defects in which one of the plaquettes has a wrong
number of neighbors with the same sign of chirality.} \label{FVortices}
\vspace*{5mm}
\end{figure}
\end{widetext}

In any ground state of the fully frustrated XY model on a honeycomb
lattice each plaquette has exactly two nearest neighbors with the same
sign of chirality and four with the opposite sign. In such a situation the
simplest idea of having a point-like excitation consists in constructing a
configuration in which one (and only one) of the plaquettes has a wrong
number of neighbors with the same sign of chirality. Indeed, such
configurations can be constructed, see Fig. \ref{FVortices}, where as in
other figures below the plaquettes with positive (negative) chiralities
are marked by the presence (absence) of a shading.

In particular, in Fig. \ref{FVortices}(a) the plaquette denoted by the
sign plus has three neighbors with the same sign of chirality (instead of
two), whereas the analogous plaquette in Fig. \ref{FVortices}(b) one
neighbor. In both configurations, in order to have the correct number of
neighbors with the same sign of chirality on all other plaquettes, three
ZEDWs have to merge at the defect.

An important property of these defects is that they are fractional
vortices with topological charges $q=\pm 1/8$. In any ground state, the
correct value of vorticity on each plaquette is
achieved  because on all four bonds separating it from the 
plaquettes with the opposite sign of vorticity variables $\theta_{jj'}$
are equal either to $+\pi/4$ or to $-\pi/4$ and sum up to $\pm \pi$. When
the number of neighbors with the same sign of vorticity is equal to one or
three (instead of two), there appears a misfit of $\pm \pi/4$ which has to
be distributed over all six bonds surrounding the corresponding plaquette.
The same misfit is also present on any closed contour surrounding the
defect independently of the size of this contour, from where it is clear
that the deviations of variables $\theta_{jj'}$ from the values they would
have in a ground state (0 or $\pm \pi/4$) decay with the distance from the
defect $R$ as $1/R$. This leads to a logarithmic divergence of the energy
of a single defect and a logarithmic (at large distances) interaction
between such defects. In accordance with their topological charges $q=\pm
1/8$, the logarithmical interaction of the fractional vortices in the
considered problem is $64$ times smaller than that of conventional
vortices with topological charges $q=\pm 1$.

The defect shown in Fig. \ref{FVortices}(c) (in which the central
plaquette has no neighbors with the same sign of vorticity) can be
considered as an overlap of two defects  of the type shown in Fig.
\ref{FVortices}(b). This statement applies both to the number of ZEDWs
which meet at the defect and to its vorticity.

\begin{figure}[t]
\includegraphics[width=80mm]{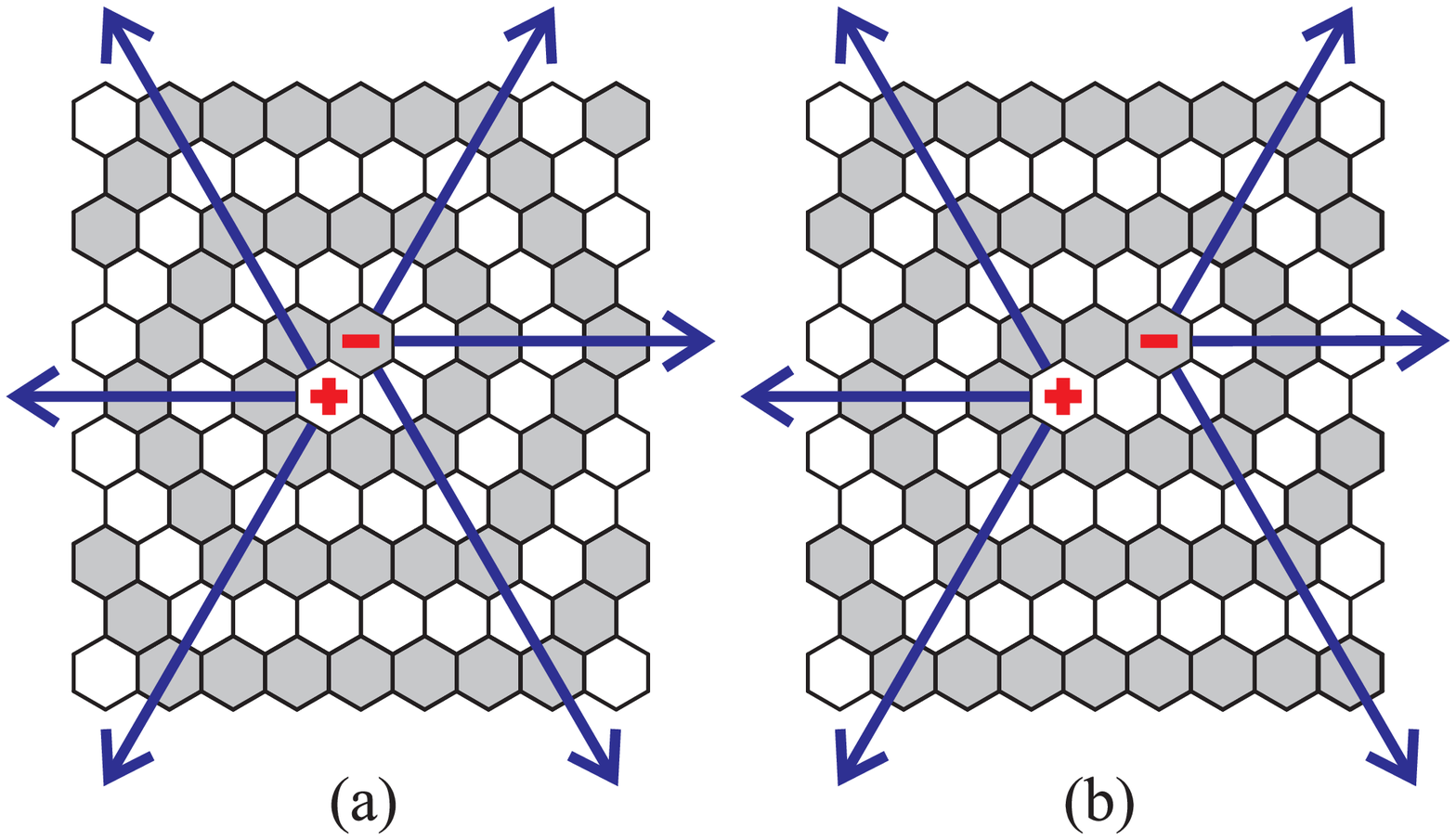}
\caption{(Color online) Neutral pairs of fractional vortices which cannot
be present in the system in the absence of free fractional vortices.}
                                                \label{Pairs-a}
\end{figure}

In Fig. \ref{FVortices} the positions of fractional vortices are denoted
by the sign plus, because in all configurations shown in this figure the
topological charge of the fractional vortex is positive. Below the
positions of fractional vortices are denoted by pluses and minuses in
accordance with the signs of their topological charges. In the case of the
Berezinskii-Villain interaction, it is possible to exactly express the
energy of any configuration of chiralities in terms of the pairwise
interaction between fractional vortices (see Appendix \ref{CG}).

\subsection{Neutral pairs of fractional vortices\label{FVP}}

Now we know how the fractional vortices look like and the next step is to
find how two fractional vortices can be combined to form a neutral pair
with a finite energy. Since each ZEDW can be ascribed a particular
direction as shown in Figs. \ref{GroundStates} and \ref{FVortices}, each
fractional vortex can be considered either as a ``sink" [the configuration
in Fig. \ref{FVortices}(a)] or as a ``source" [the configurations in Figs.
\ref{FVortices}(b) and \ref{FVortices}(c)] of domain walls.

Apparently in the case of periodic boundary conditions the number of
sources must be equal to the number of sinks [naturally, for the correct
balance the configurations of the type shown in \ref{FVortices}(c) have to
be counted as double sources]. It is impossible to construct an isolated
pair of fractional vortices from two sink configurations because the
domain walls meeting in them would have to intersect each other, which
would imply the existence of other defects close by. For the same reason
it is impossible to construct an isolated pair from two double sources. On
the other hand, although locally a neutral pair consisting of two sources
looks like a legitimate object (see Fig. \ref{Pairs-a}), such pairs cannot
appear because in the absence of unpaired fractional vortices there will
be no sinks to compensate for these sources [in that respect they resemble
the configuration shown in Fig. \ref{GroundStates}(d)].

\begin{widetext}

\begin{figure}[h]
\includegraphics[width=130mm]{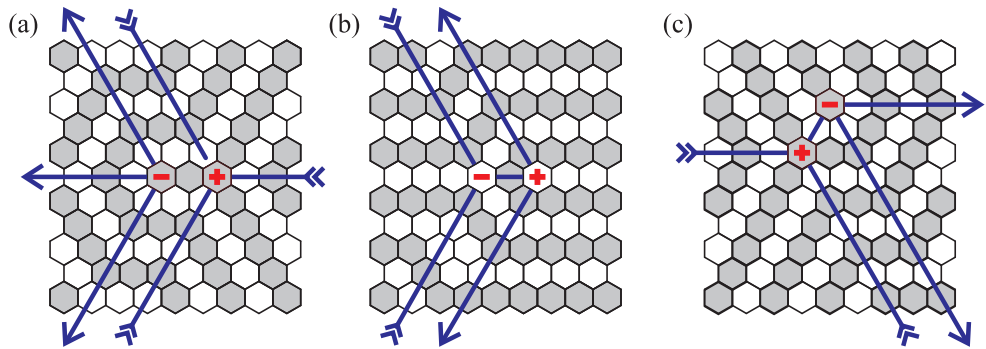}
\caption{(Color online) Neutral pairs of fractional vortices which can
exist in the absence of free fractional vortices.}
                                                \label{Pairs-b}
\end{figure}

\end{widetext}

Therefore, in a domain wall network formed by small pairs of fractional
vortices, each pair should consist of a source and a sink. Such pairs can
be divided into three classes, exemplified in Fig. \ref{Pairs-b}. The
first of them is an intersection of a single ZEDW and a double domain
wall, that is, a pair of parallel ZEDWs (which change their orientation
after crossing a single wall). An example of such a pair of fractional
vortices is shown in Fig. \ref{Pairs-b}(a).

The second type of the fractional-vortex pairs  are the bends on a double
domain wall, see Fig. \ref{Pairs-b}(b). For such a pair to be neutral the
distance between the two single walls forming the double wall has to be
even (in the considered model it is natural to measure the distance
between domain walls in units of $h_\triangle$, 
the height of the triangular cell of the dual lattice). If the distance
between two parallel ZEDWs would be odd (for example, the minimal
distance), in the configuration analogous to the one shown in Fig.
\ref{Pairs-b}(b) both fractional vortices would be of the same sign.

The third class of the fractional-vortex pairs  corresponds to a double
domain wall ending on a single domain wall, see Fig. \ref{Pairs-b}(c). In
this case the distance between the two parallel ZEDWs also has to be even
in order to avoid a situation when both fractional vortices in a pair are
of the same sign.

\section{ The lowest-temperature phase transition \label{first}}

When a linear defect (for example, a domain wall) has proper energy or
free energy per unit length $\epsilon$ and additionally allows for the
creation of point-like defects with energy $E_{\rm D}$, its total free
energy per unit length is given by
\vspace*{-3mm}
\begin{equation}                                               \label{FLD}
        F_{\rm} = \epsilon-cT\ln\left(1+e^{-E_{\rm D}/T}\right)
              \approx \epsilon-cT e^{-E_{\rm D}/T}\,,
\end{equation}
where $c$ is the number of places per unit length available for the
formation of a point-like defect and we have assumed that $T\ll E_{\rm
D}$. A natural choice for the unit length when speaking about domain walls
in a lattice model is the lattice constant either of the direct or of the
dual lattice, depending on whether the domain walls can be associated with
the bonds of the former or of the latter. Then $c$ has to be equal either
to one or to some simple fraction (or integer) of the order of one. In our
system, it is convenient to measure the length of domain walls in lattice
constants of the dual (triangular) lattice, $a_\triangle$.

It follows from Eq. (\ref{FLD}) that for $\epsilon\ll E_{\rm D}$ the
temperature $T_{\rm *}$ at which $F_{\rm }$ changes sign lies between
$\epsilon/c$ and $E_{\rm D}$ ($\epsilon/c\ll T_*\ll E_{\rm D}$) but
depends on $E_{\rm D}$ much stronger than on $\epsilon$:
\begin{equation}
\label{TDW}
    T_{\rm *}= \frac{E_{\rm D}}{\ln(cT_{\rm *}/\epsilon)}\;.
\end{equation}
It is clear from Eq. (\ref{TDW}) that the first linear defects whose free
energy vanishes with the increase in temperature are those that allow for
the formation of point-like defects with the lowest energy.

It follows from the analysis of Sec. \ref{FVP} that the lowest-energy
point-like excitation on a linear defect in the considered problem is a
bend on a doubly-spaced double domain wall (DDDW), consisting of two
single domain walls separated by distance $2h_\triangle$, see Fig.
\ref{Pairs-b}(b). A numerical minimization of energy in a finite system
(complemented with the extrapolation to the infinite size) shows that in
the model with $V(\theta)=-J\cos\theta$ the energy of this defect $E_{\rm
B}$ is one order of magnitude smaller than the coupling constant, $E_{\rm
B}/J\approx 0.111$. Taking for DDDW $\epsilon=2f_{\rm DW}$ with $f_{\rm
DW}$ given by Eq. (\ref{fDW}), $c=1$ and $E_{\rm D}=E_{\rm B}$, one
obtains from Eq. (\ref{TDW}) that $T_{\rm DDDW}$, the temperature where
the free energy of a DDDW vanishes, is about $14$ times lower than $E_{\rm
B}$ and two orders of magnitude lower than the coupling constant,
\begin{equation}
T_{\rm DDDW}\approx 0.81\times 10^{-2}\,J\,.
\end{equation}

\begin{widetext}

\begin{figure}[bp]
\vspace*{-4mm}
\includegraphics[width=140mm]{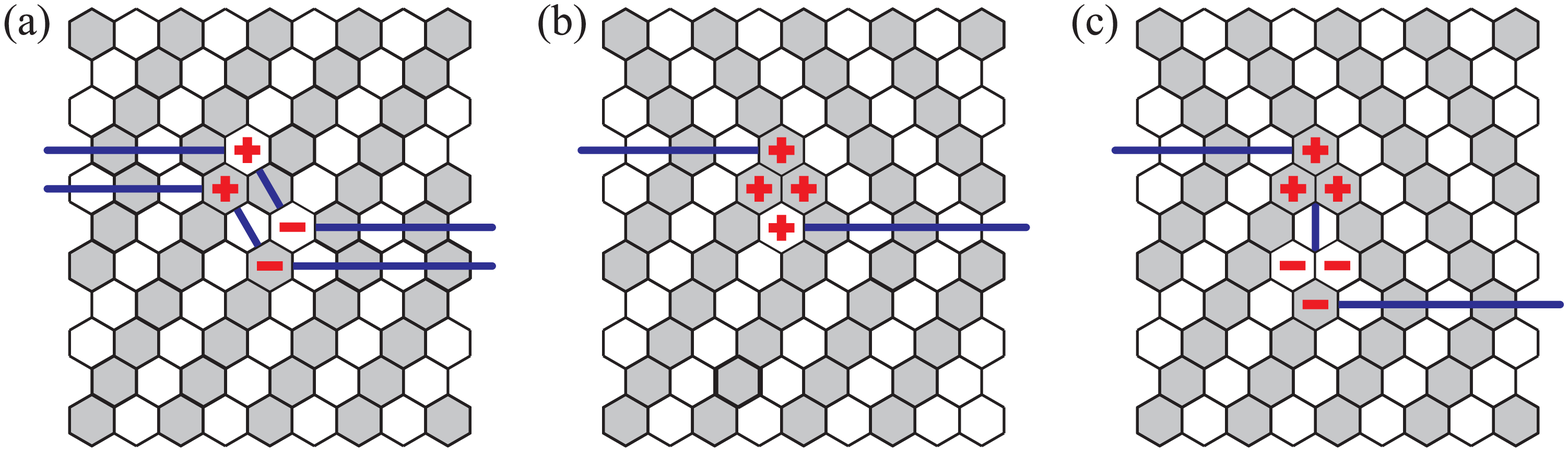} \caption{(Color online)
Kinks on a singly-spaced double domain wall (a) and a single domain wall
[(b) and (c)].}
                                                \label{Kinks}
\end{figure}
\end{widetext}

\vspace*{-3mm} In the vicinity of $T_{\rm DDDW}$, the influence of the
point-like defects on the free energy of other types of domain walls can
be neglected. This follows from the analysis of their energies. In
particular, a bend on a double domain wall with the minimal separation
between the walls, $h_\triangle$ (a singly-spaced double domain wall),
contains two fractional vortices with the same sign of topological charge
$q=\pm \frac{1}{8}$. Therefore, its energy is logarithmically divergent
and the lowest-energy excitation on such a wall is not a bend but a kink
formed by two bends, see Fig. \ref{Kinks}(a). Since this object can be
considered as a neutral pair of fractional vortices with topological
charges $q=\pm \frac{1}{4}$, its energy $E_{\rm K}$ has to be larger than
that of a bend on a DDDW (a neutral pair of fractional vortices with
topological charges $q=\pm \frac{1}{8}$). This is confirmed by a numerical
calculation revealing that $E_{\rm K}\approx 0.287\, J$, that is, the
energy of such a kink is 2.6 times larger than that of a bend on a DDDW.
This ensures that at $T\approx T_{\rm DDDW}$ the 
correction to the free energy of a singly-spaced double domain wall is
nine orders of magnitude smaller than its bare value and therefore can be
neglected.

A single domain wall even cannot make a bend, because the direction of
such a wall is uniquely determined by the directions of vortex stripes in
the states which it separates. The simplest point-like excitation on a
single domain wall is a kink with height $2h_\triangle$. Such a kink is
formed by four fractional vortices with $q=\pm 1/8$ having the same sign
[see Fig. \ref{Kinks}(b)] and therefore has topological charge $q = \pm
\frac{1}{2}$. Accordingly, the minimal-energy excitation on a single
domain wall is a neutral pair of such kinks forming a kink of
height $4h_\triangle$, see Fig. \ref{Kinks}(c). 
It is clear that the energy of this complex object $E^{\rm s}_{\rm K}$ has
to be about one order of magnitude larger than that of a bend on
\makebox{a DDDW (for the case of the Berezinskii-Villain interac-}

\hspace*{-3.5mm}tion, the values of $E_{\rm B}$, $E_{\rm K}$, $E^{\rm
s}_{\rm K}$ and other finite-energy defects mentioned in the text can be
found in Appendix \ref{CG}).

\begin{figure}[bp]
\includegraphics[width=80mm]{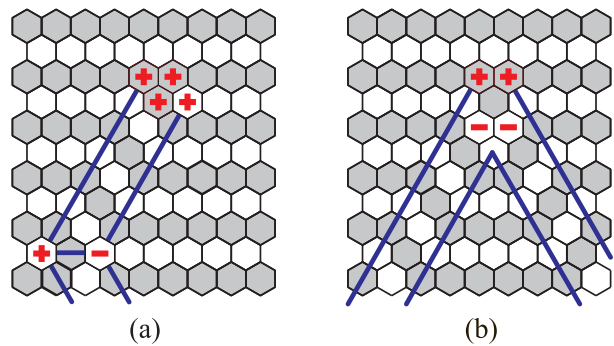}
\caption{(Color online) (a) an end-point of a DDDW is a $q=\pm\frac{1}{2}$
vortex; (b) an end-point of a finite size defect formed by two fluctuating
DDDW. } \label{EndPoints}
\end{figure}

From this one can conclude that in the vicinity of $T_{\rm DDDW}$ all
other topological excitations save DDDWs can be neglected. Note that since
the vortex patterns on the two sides of a DDDW are the same, a DDDW is not
a topologically stable object and in principle can end somewhere, see
Fig. \ref{EndPoints}(a). This is especially evident when one notices that
any DDDW is nothing else but a line of plaquettes with alternating
chiralities on which the signs of all chiralities are reversed. However,
the states on the two sides of a DDDW are not exactly the same, namely,
they differ by a phase rotation by $\pi$. In order to compensate for this
misfit an end-point of a DDDW has to be a fractional vortex with
topological charge $q=\pm \frac{1}{2}$. The same conclusion also follows
from observing that an end-point is formed by four fractional vortices
with $q=\pm \frac{1}{8}$ having the same sign, see Fig.
\ref{EndPoints}(a).

At temperatures we are discussing, vortices with topological charges
$q=\pm \frac{1}{2}$ have to be bound in small neutral pairs. Therefore, in
the nearest vicinity of any end-point there always has to be present
another end-point with the opposite topological charge, so that the two
DDDWs can be considered as a continuation of one another [analogous
situation is known to exist in the unfrustrated XY model in which the
interaction function $V(\theta)$ in addition to the main minimum at
$\theta=0$ has an additional minimum at $\theta=\pi$ with almost the same
depth \cite{K85}]. The orientation of a fluctuating DDDW is determined by
the orientation of stripes in the vortex pattern on its sides, on the
average such a wall will be perpendicular to the direction of stripes.

At $T>T_{\rm DDDW}$ the free energy of a single DDDW becomes negative
which suggests that there should appear a finite concentration of such
walls $\rho$ restricted by their repulsion. Since this repulsion is of
contact nature, the dependence of $\rho$ on the distance from the critical
temperature can be expected to be of the Pokrovsky-Talapov \cite{PT80}
type, $\rho\propto (T-T_{\rm DDDW})^{1/2}$.

However, in addition to the formation of infinite DDDWs the system allows
also for the creation and annihilation of pairs of fluctuating DDDWs, see
Fig. \ref{EndPoints}(b). This means that in addition to infinite DDDWs
crossing the whole system, there should be present finite-size defects
formed by two fluctuating DDDWs, which on both sides of the defect merge
together as shown in Fig. \ref{EndPoints}(b). In the theory of the
commensurate-incommensurate transitions the objects where $n$ solitons
merge together are known as dislocations and the case of $n=2$ corresponds
to the double degeneracy of the ground state. \cite{BPT} In our system,
the presence of a two-fold degeneracy manifests itself through the fact
that the states on the two sides of a DDDW differ from each other by phase
rotation by $\pi$. As a consequence of this, after crossing two DDDWs one
returns to the same state as before. Accordingly, the phase transition
related to the proliferation of DDDWs has to be of the Ising type.

Since the states on the two sides of a DDDW correspond to the same vortex
pattern but  differ by a phase rotation by $\pi$, the phase transition
related to their proliferation does not destroy the long-range order in
terms of chirality but leads to a partial suppression of phase
correlations. Namely, an algebraic decay of correlation function $C_1({\bf
r})$ is replaced by by an exponential one, whereas the correlations of the
double phase [described by $C_2({\bf r})$] remain algebraic. In the
Pokrovsky-Talapov regime, the temperature dependence of the correlation
radii describing the decay of $C_1({\bf r})$ is characterized by two
different values of the exponent $\nu$ ($\nu=1/2$ and $\nu=1$) for the two
directions, whereas in the Ising regime the value of $\nu=1$ is the same
for both directions.

The crossover from the Pokrovsky-Talapov behavior to the Ising one must
take place very close to the phase transition temperature, because each
point where two DDDWs are created or annihilate is a pair of fractional
vortices with topological charges $q=\pm \frac{1}{4}$ and accordingly has
larger energy than a bend on DDDW. It should be emphasized that when the
temperature is about 14 times smaller than $E_{\rm B}$ even a relatively
small increase in energy in comparison with $E_{\rm B}$ leads to the
suppression of the Boltzmann factor $\exp(-E_{\rm D}/T)$ by orders of
magnitude.

Just above $T_{\rm DDDW}$ the distance between the DDDWs $L$ is much much
larger then their ``effective width" $\xi$ given by the average distance
between neighboring bends on a wall, $\xi \equiv \xi(T) = \exp(E_{\rm
B}/T)$. Since it is evidently disadvantageous to have $L\ll \xi$ (this
would force the concentration of bends with positive energy 
to be much larger than the optimal one), the ratio $L/\xi$ has to saturate
with the increase in temperature at $L/\xi\sim 1$. For this to take place,
the Boltzmann factor
\begin{equation}
w_{\rm B}(T)\equiv\exp(-E_{\rm B}/T)
\end{equation}
has to become much larger than $f_{\rm DW}/T$, which allows one to neglect
the first term in Eq. (\ref{FLD}). Due to the exponential dependence of
$w_{\rm B}(T)$ on $T$, this is achieved already in a close vicinity of
$T_{\rm DDDW}$.

In this regime the free energy of the fluctuating DDDWs (per lattice
plaquette) is of the order of \makebox{$-T\exp(-2E_{\rm B}/T)$}, where one
factor $\exp(-E_{\rm B}/T)$ is directly related to the energy of the
elementary point-like defect and the other one appears because the
distance between neighboring DDDWs has to be exponentially large and,
accordingly, the number of places where these point-like defects can be
created is exponentially suppressed. However, in the situation when the
fluctuation-induced free energy of domain walls can be neglected (or is
absent from the beginning, as in the case of the Berezinskii-Villain
interaction), one has to consider also other possibilities for the
appearance of point-like defects in the system. This task is carried out
in the next section.

\section{Uniaxial network state}

In this section we first discuss what happens at low temperatures if
$f_{\rm DW}$, the free energy of straight domain walls per unit length,
is exactly equal to zero and therefore does not induce a removal of the
accidental degeneracy of the ground states. And after finding an answer to
this question, we return to the problem with small but finite $f_{\rm
DW}>0$.

For $f_{\rm DW}=0$,  one can expect that at low temperatures a typical
configuration in addition to having parallel straight domain walls may
include a small concentration of local defects (with finite energies) at
which these domain walls change their orientations. However, a finite
concentration of such objects can appear at arbitrary low temperature only
if the entropy (per defect) of the network of domain walls connecting them
can be made arbitrary large by the decrease of their concentration, as in
the gas of non-interacting particles. Without a detailed analysis it is
not evident what structure such a network can have and whether its
appearance induces a complete disordering of the vortex pattern.

In Sec. \ref{PointLikeDefects} we have shown that the lowest-energy local
defects in our system are pairs of fractional vortices with topological
charges $q=\pm \frac{1}{8}$. Four different classes of such defects are
illustrated in Figs. \ref{Pairs-a} and \ref{Pairs-b}. However, a network
of domain walls cannot contain the pairs of the type shown in Figs.
\ref{Pairs-a}  because the domain walls ``emitted" by them can end only on
free (that is, unpaired) fractional vortices, which at low temperatures
cannot be present in the system.

\begin{widetext}

\begin{figure}[h]
\includegraphics[width=120mm]{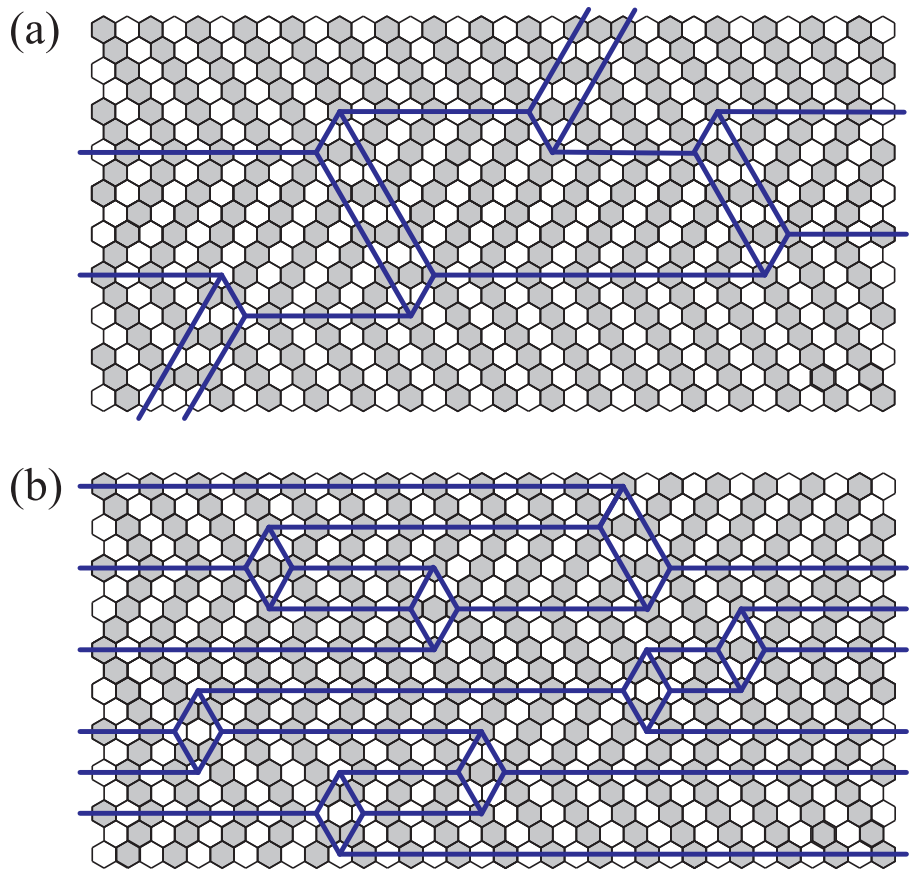}
\caption{(Color online) A network formed by parallel single domain walls
connected by double-spaced double domain walls: \\ (a) with relatively
small density $n_{\rm DW}$ of single walls\,; (b) with $n_{\rm DW}\approx
1/2$.}
                                                \label{Network}
\end{figure}
\newpage
\end{widetext}

The neutral pair of fractional vortices shown in Fig. \ref{Pairs-b}(a) is
an intersection of single and double domain walls. In order to have a
finite concentration of the defects of such a kind, it is necessary to
have a sequence of parallel single domain walls and a sequence of double
domain walls crossing them. However, the entropy of such a network will be
proportional to the number of the walls in these sequences, which does not
give the entropic contribution to free energy a chance to overcome the
positive term related to the proper energy of the defects and proportional
to the number of the intersections. This suggests that such defects cannot
play a substantial role in the spontaneous formation of a domain-wall
network at the lowest temperatures.

In contrast to that, the neutral fractional-vortex pairs of the two other
types shown in Fig. \ref{Pairs-b}(b) and Fig. \ref{Pairs-b}(c) allow for
having an arbitrary large entropy per defect. In particular, the
configuration shown in Fig. \ref{Pairs-b}(b) is nothing else but a bend on
a double-spaced double domain wall. In the last paragraph of Sec.
\ref{first} we have argued that for $f_{\rm DW}\rightarrow 0$ the free
energy (per site) of the sequence of such double walls has to be of the
order of $-T\exp(-2E_{\rm B}/T)$, where $E_{\rm B}$ is the energy of a
single bend. In this estimate, the exponent contains not just the ratio
$E_{\rm B}/T$ like it would be if different point-like defects could be
created independently of each other, but $2E_{\rm B}/T$. The second factor
$\exp(-E_{\rm B}/T)$ appears because double domain walls have to be
separated by distances of the order of $\xi\equiv\exp(E_{\rm B}/T)$, and
therefore the number of places where the point-like defects with energy
$E_{\rm B}$ can appear is exponentially suppressed.

Another possibility to get a free energy of comparable order
consists in considering configurations in which the energy of point-like
defects is larger than $E_{\rm B}$, but the number of places available for
their creation is just proportional to the area of the system without an
additional exponential suppression. This happens when one considers a
configuration with a finite density $\rho_\|$ of parallel single domain
walls and inserts into it a finite concentration of point-like defects
each of which is a double-spaced double domain wall starting on one single
wall and ending on a neighboring one, as shown in Fig. \ref{Network}(a).
The energy of these defects is close to $2E_{\rm B}$
and, accordingly, 
for $\rho_\|\ll 1$ the free energy of such a ``uniaxial" network can be
estimated as
\begin{equation}                                            \label{Fnw}
F_{\rm UNW}\sim -\rho_\|\,T\exp(-2E_{\rm B}/T)\,.
\end{equation}
Note that Fig. \ref{Network} gives just a schematic representation of the
structure of corresponding states. In reality, at $T\ll E_{\rm B}$ the
typical distances between neighboring local defects have to be
exponentially large. Accordingly, the mutual influence of the local
defects can be neglected.

It is clear from Eq. (\ref{Fnw}) that it is more profitable to have
$\rho_\|$ of the order of one rather than $\rho_\| \ll 1$. The
concentrations close to 1/2 (the average concentration of parallel domain
walls in a typical ground state) seem to be the most optimal because they
optimize both the number of the available configurations as well as the
energy of typical local defects.

\begin{figure}[bp]
\includegraphics[width=80mm]{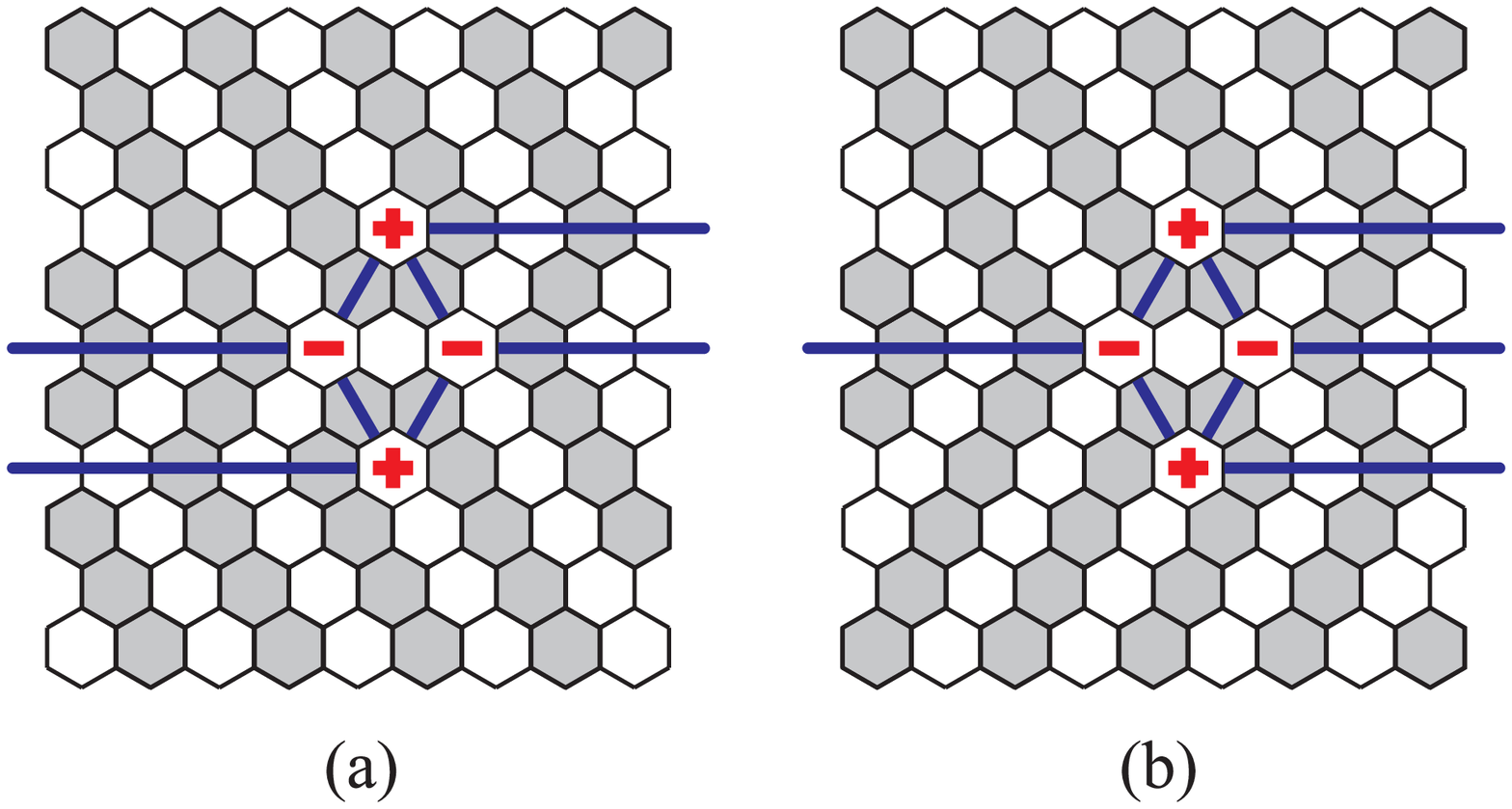}
\caption{(Color online) The structure of the typical local defects
participating in the formation of the uniaxial network state [shown in
Fig. \ref{Network}(b)]. }
                                                \label{Rhombi}
\end{figure}

The minimum of energy of two fractional-vortex pairs  connected by a
double-spaced double domain wall is achieved when the four fractional
vortices form a symmetric rombus, as shown in Fig. \ref{Rhombi}(a). Both
for \makebox{$V(\theta)=-J\cos\theta$} and for the Berezinskii-Villain
interaction, the energy of such a defect, $E_{\rm R}$ is smaller than
$2E_{\rm B}$ by approximately $11\%$. The same is true for another defect
with the same rhombic arrangement of four fractional vortices shown in
Fig. \ref{Rhombi}(b). Accordingly the free energy of the domain wall
network containing such defects (and maybe some other defects as well)
at the lowest temperatures will be of the order of $-T\exp(-E_{\rm R}/T)$,
that is, lower than the free energy of the sequence of fluctuating double
walls.

\begin{figure}[bp]
\includegraphics[width=80mm]{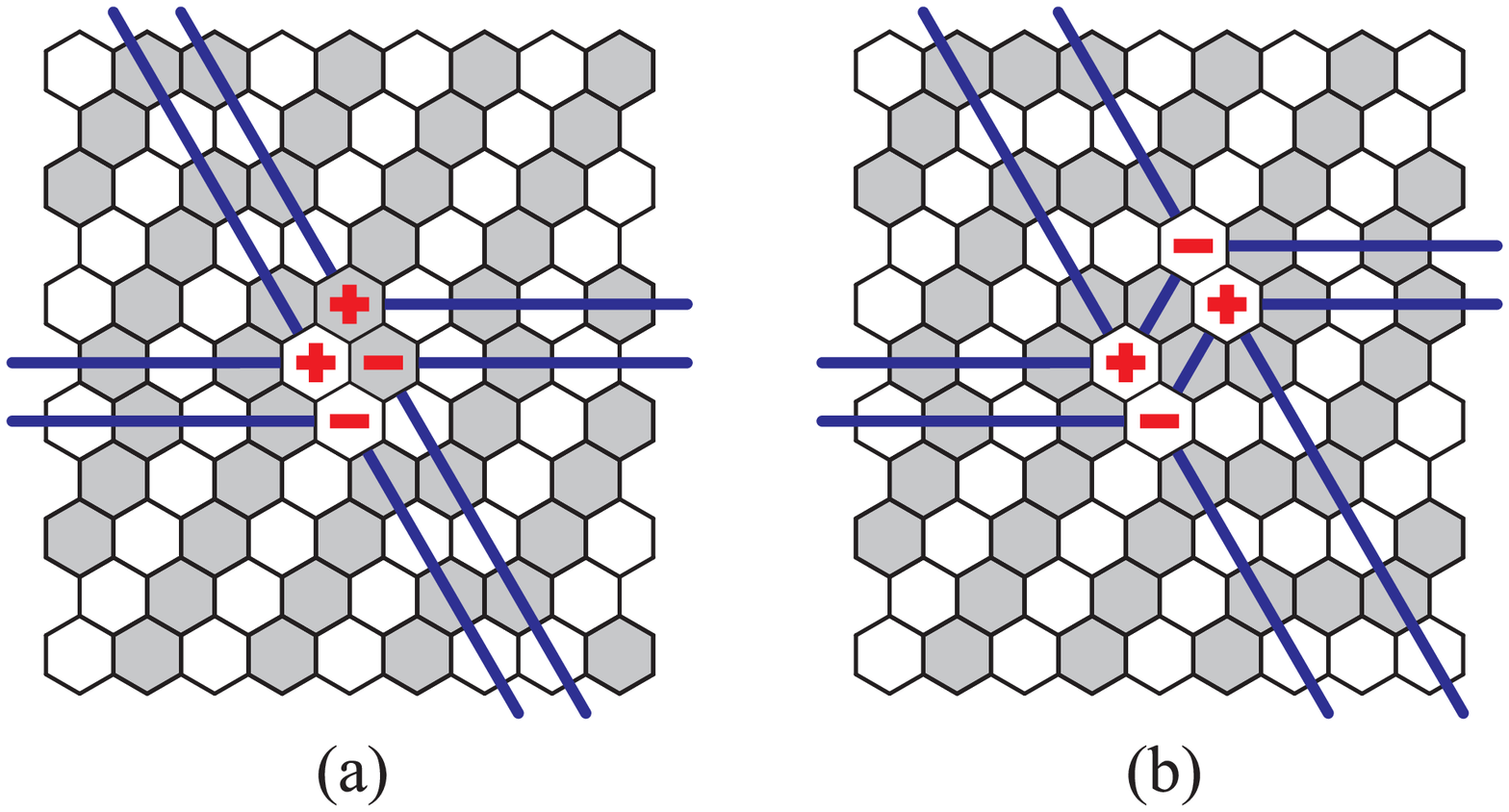}
\caption{(Color online) Intersections of double domain walls.}
                                                \label{Intersections}
\end{figure}

Note that the list of local defects whose energy is lower than $E_{\rm R}$
in addition to configurations which cannot participate in the formation of
a domain-wall network [Fig. \ref{Pairs-a}] and bends on double-spaced
double domain walls [Fig. \ref{Pairs-b}(b)] includes only the
intersections of a double domain wall either with a single wall [Fig.
\ref{Pairs-b}(a) ] or another double wall [Fig. \ref{Intersections}]. The
presence of such local defects is not sufficient for the construction of a
domain wall network whose entropy dominates over the energy of the
defects. This ensures that for $f_{\rm DW}=0$, the main role in the
formation of the domain-wall network is played by the rhombic defects
[shown in Fig. \ref{Rhombi}] whose presence leads to the formation of  the
uniaxial network state schematically depicted in Fig. \ref{Network}(b).

Reminder that in a typical ground state the concentration of domain walls
parallel to each other is equal to $1/2$. For $f_{\rm DW}=0$ at the lowest
temperatures the system will be in the same phase, the main difference
consisting in the presence of an exponentially small concentration of
local defects, which shift the positions of domain walls but do not lead
to the change of their orientation, see Fig. \ref{Network}(b). Like at
$T=0$, the presence of a finite concentration of parallel domain walls
leads to the intermixing of four out of six striped vortex patterns.
However, there remains a triple degeneracy related to the orientation of
domain walls. The order parameter suitable for describing this kind of
ordering has been introduced in Sec. \ref{T=0-LRO}.

In terms of phase correlations, the main qualitative difference with the
case of \makebox{$T=0$} is that at $T>0$ correlation function $C_p({\bf
r})$ with $p=8$ acquires an algebraic decay (analogous correlation
functions with $p<8$ decay exponentially even at $T=0$). However, this
property is related not to the appearance of local defects in the
domain-wall network but to the presence of spin waves.

Thus, we have found that the formation of local defects works as a
mechanism for the removal of the accidental degeneracy of the ground
states which makes the free energy of the system dependent on the
concentration of parallel single domain walls $\rho_\|$, with
$F\left(\rho_{\|}\right)$ having the minimum in the vicinity of
$\rho_\|=1/2$ (in the limit of $T\to 0$ exactly at $\rho_\|=1/2)$. Now we
have to remember that in the absence of this mechanism $F(\rho_\|)$ is
equal at small $\rho_\|$ to $ f_{\rm DW}(T)\rho_\|$ and, accordingly, is
minimal at $\rho_\|=0$. When both mechanisms are taken into account, the
function $F\left(\rho_{\|}\right)$ at least in some interval of
temperatures will have two minima whose depths change with temperature. At
the temperature where these two depths become equal, a first-order phase
transition has to take place.

For $f_{\rm DW}>0$, a possibility to describe the system assuming $f_{\rm
DW}=0$ exists if the ratio
\begin{equation}                                             \label{kappa}
\kappa(T)=\frac{f_{\rm DW}(T)}{T\exp(-E_{\rm R}/T)}
\end{equation}
is much smaller then one. For the model with
\makebox{$V(\theta)=-J\cos\theta$} the substitution of the expression
(\ref{fDW}) for $f_{\rm DW}(T)$ into Eq. (\ref{kappa}) gives that at
$T/J=0.02$ the value of $\kappa(T)$ is close to $0.03$. Accordingly, at
$T/J\approx0.02$ the corrections to free energy related to the
positiveness of $f_{\rm DW}$ can be neglected and one can expect that the
free energy of the system is minimal in the uniaxial network state. On the
other hand, at $T/J=0.015$ the negative contribution to the free energy
related to the appearance of the point-like defects with energy $E_{\rm
R}$ cannot overcome the positive term related to the proper free energy of
domain walls and any reasons for the appearance of a uniaxial network with
a large concentration of parallel domain walls are absent.

From this it is clear that the first-order transition mentioned above
takes place at some temperature between $0.015\,J$ and $0.02\,J$. Below
this transition the minimum of free energy is achieved in the phase with a
sequence of fluctuating double walls described in Sec. \ref{first} and
above it in the uniaxial network with a relatively high density of
parallel domain walls. 
The conclusion that this phase transition is of the first order is
additionally confirmed by checking that the presence of fluctuating
double-spaced double domain walls does not lead to a decrease of free
energy of a single domain wall crossing them. The advantages of the
uniaxial domain-wall network manifest themselves only when the density of
single domain walls is of the order of $1/2$. They consist in the
possibility of having local defects of an additional type shown in Fig.
\ref{Rhombi}, which increases the number of configurations with the given
number of the defects, that is with the given energy.

\section{Phase transition to a disordered phase}

For the complete disordering of vortex pattern the network of domain walls
has to  have a structure leading to an unbiased representation of all six
striped patterns. If all fractional vortices are bound in small neutral
pairs this is impossible. The existence of a finite concentration of such
pairs as those shown in Fig. \ref{Pairs-a} is prohibited by a simple
conservation law explained in Sec. \ref{FVP}, whereas the presence of
small concentration of pairs of the three types represented in Fig.
\ref{Pairs-b} leads to the intermixing of only four striped states out of
six. Such a uniaxial network has a preferable orientation of domain walls.

In order to have an equal representation of all six striped vortex
patterns there should exist a finite concentration of free fractional
vortices which are not bound in pairs. If their positions could be
arbitrary (that is, not restricted by the requirement for them to be
connected by domain walls) free fractional vortices would appear when the
two logarithmically divergent contributions to the free energy of a single
fractional vortex
\begin{equation}
F_{\rm FV}=E_{\rm FV}-TS_{\rm FV}
\end{equation}
compensate each other. Here
\begin{equation}                                               \label{E_FV}
E_{\rm FV}\approx E_0\ln L_{\rm max}\,,~~~ E_0=
\left(\frac{1}{8}\right)^2\pi\Gamma
\end{equation}
is the energy of a fractional vortex and $S_{\rm FV}\approx 2\ln L_{\rm
max}$ its entropy, whereas $\Gamma\equiv\Gamma(T)$ is the helicity modulus
\cite{comm-anisotropy} and $L_{\rm max}\to\infty$ the size of the system.
The prelogarithmic factor in  $F_{\rm FV}$ becomes equal to zero at
\makebox{$T=\frac{1}{2}E_0=\frac{\pi}{128}\Gamma$.} On a honeycomb lattice
with the conventional or the Berezinskii-Villain interaction $\Gamma(T=0)$
is respectively slightly below or slightly above $J/2$. Substitution of
this value suggests that in the absence of any restrictions related to
domain walls free fractional vortices would appear at $T_{\rm FV}\approx
0.01 J$. \cite{renorm}

\begin{figure}[bp]
\includegraphics[width=75mm]{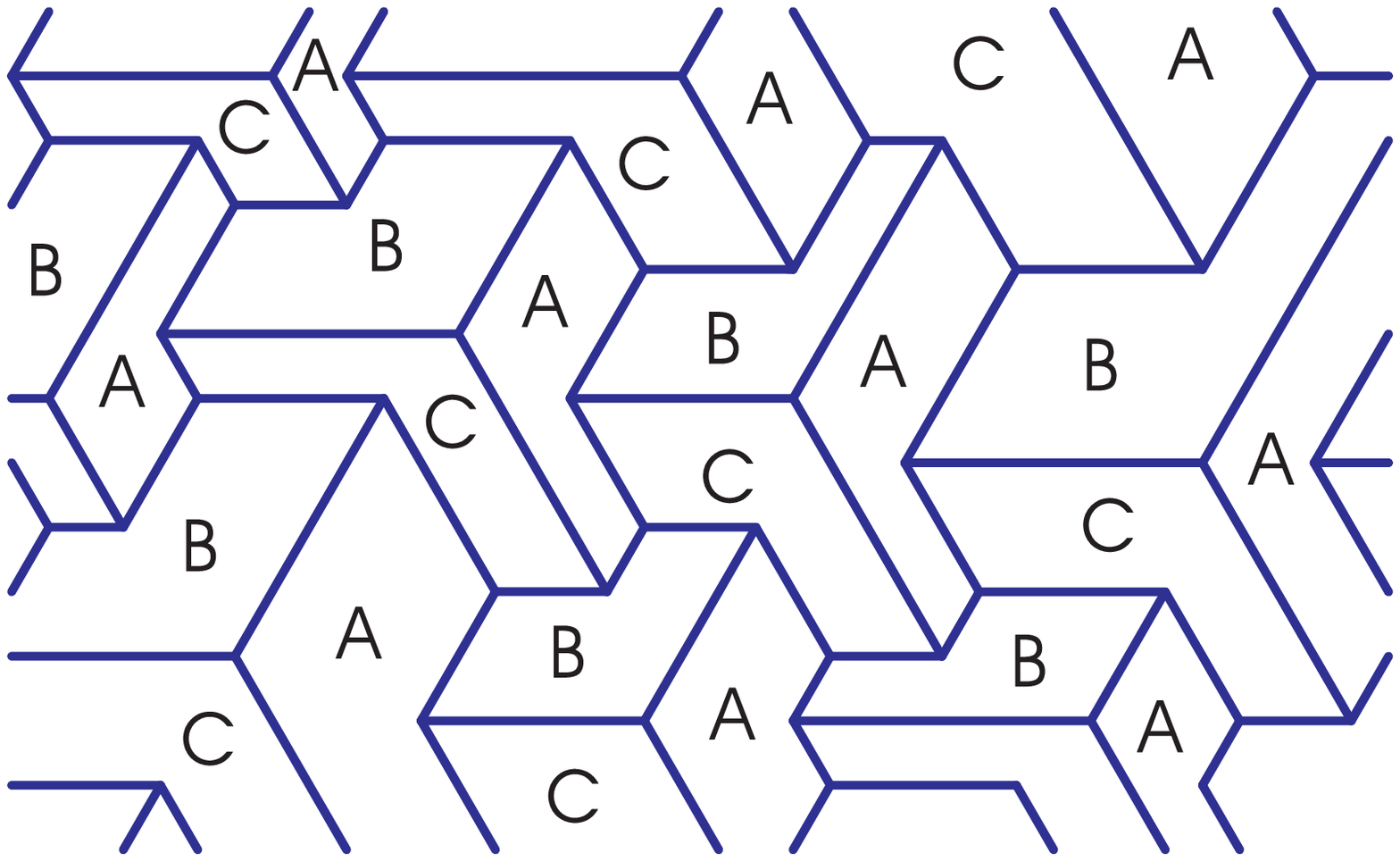}
\caption{The structure of a domain wall network in which all six regular
vortex pattern are intermixed. }
                                                \label{IsingLikeNetwork}
\end{figure}

Naturally, in the fully frustrated XY model on a honeycomb lattice the
entropy of the system of fractional vortices is substantially decreased by
the requirement that they have to be connected by straight domain walls
(with maybe some rare kinks on them). Moreover, the direction of each
domain wall is uniquely determined by the directions of the vortex stripes
in the two states which it separates. Nonetheless, these rather strong
restrictions still allow for constructing a domain-wall network which
leads to an unbiased representation of all six striped vortex patterns. A
possible structure of a such network is schematically shown in Fig.
\ref{IsingLikeNetwork}. Here the letters A, B and C are used to denote the
domains with three different orientations of stripes. Note that all walls
between A and B are parallel to each other. The same is true for all walls
between B and C, as well as for all walls between C and A. For the sake of
lucidity Fig. \ref{IsingLikeNetwork} does not show  which of the two
versions of A, of B, or of C (related to the change of the signs of all
chiralities) is realized in each particular domain. This depends on the
exact positions of domain walls.

Quite remarkably, the entropy (per node) of such a network logarithmically
depends on typical distance between neighboring nodes \cite{K05} like it
would be in a gas of free particles. This is so because each domain of a
network with such a structure can be shifted in parallel to six domain
walls which end up in its corners without changing the number of domains
or the total length of the domain walls. Thus, if the typical distance
between neighboring nodes of the network is of the order of $L \gg 1$,
then each domain can occupy a number of positions which is proportional to
$L$. Since the number of domains is equal to one half of the number of
nodes, the domain-wall-network entropy per fractional vortex is given (for
large enough $L$) by $\frac{1}{2}\ln L$ per fractional vortex. This
quantity is only four times smaller than in the absence of restrictions
induced by the presence of domain walls.

The logarithmic behavior of the domain-wall network entropy was first
discovered by Villain \cite{Vil80a} when considering a honeycomb network
in which each domain has the shape of a hexagon with all angles equal to
120 degrees.
Such a network is formed in a system with a threefold degeneracy in which
a domain wall of a given type (for example, a wall between states A and B)
can have only three particular orientations out of six that seem to be
possible when one looks just at the symmetry of the lattice. A domain-wall
network with the structure shown in Fig. \ref{IsingLikeNetwork} (in which
each domain also has the shape of a hexagon but with angles of 60, 120 and
240 degrees), has been proposed for the Ising model on a triangular
lattice with the antiferromagnetic interaction of the nearest, second and
third neighbors. \cite{K05} For some relation between parameters, the
ground states of this Ising model have the striped structure \cite{Metc}
completely analogous to that of the striped ground states of the
considered XY model. However, the important difference between the two
models is that in the model we discuss now the nodes of the domain wall
network have a long-range logarithmical interaction.

In accordance with that, in the fully frustrated XY model on a honeycomb
lattice the free energy (per lattice plaquette) of a domain wall network
with the structure shown in Fig. \ref{IsingLikeNetwork}
is given by
\begin{equation}                                               \label{F_NW}
    F_{\rm NW}=\frac{c_1}{L^2}\left(E_{0}-\frac{T}{2}\right)\ln L
    +F_{\rm {NW}}^{(1)}(L)\,,
\end{equation}
where $c_1\sim 1$ is a constant of purely geometrical origin, whereas
$L\gg 1$ is the typical distance between neighboring fractional vortices.
The last term in Eq. (\ref{F_NW}) describes the dominant corrections to
the logarithmic terms and for $L\gg 1$ can be estimated as $F_{\rm
NW}^{(1)}(L)\sim E_0/L^2$.

At $T<T_{\rm FV}=2E_{0}$, when both terms in Eq. (\ref{F_NW}) are positive
the minimum of $F_{\rm NW}$ is achieved when $L\to\infty$, that is, in the
absence of a domain-wall network. On the other hand, at \makebox{$T>T_{\rm
FV}$} the free energy of the system at large $L$ decreases with the
decrease in $L$. Therefore, the optimal value of $L$ at such temperatures
is determined by the interplay between the two terms in Eq. (\ref{F_NW}).
It is then clear without performing any calculations, that when $T$ tends
to $T_{\rm FV}$ from above, the value of $L$ minimizing $F_{\rm NW}$ goes
to infinity.

However, all the conclusions of the previous paragraph would be applicable
only if any other fluctuations save the formation of a domain wall network
with the structure shown in Fig. \ref{IsingLikeNetwork} are prohibited. In
reality, one expects that at such temperatures
the system is in the uniaxial-network state with a large concentration of
parallel domain walls. In this situation, the structure of the domain-wall
network related to the appearance of unpaired fractional vortices also can
be illustrated by Fig. \ref{IsingLikeNetwork}, where now the domains
denoted by the same letter correspond to uniaxial-network states with the
same dominant orientation of domain walls. Nonetheless, exactly as before,
each node of the network corresponds to an unpaired fractional vortex with
topological charge $q=\pm \frac{1}{8}$, which means that one can still use
expression (\ref{F_NW}) for describing the energy of the interaction of
unpaired fractional vortices plus the entropy of the network related to
fluctuations of positions of different domains. However, one also has to
add to this expression the energy of the boundaries between different
uniaxial networks.

\begin{figure}[bp]                                 \label{Boundary}
\includegraphics[width=70mm]{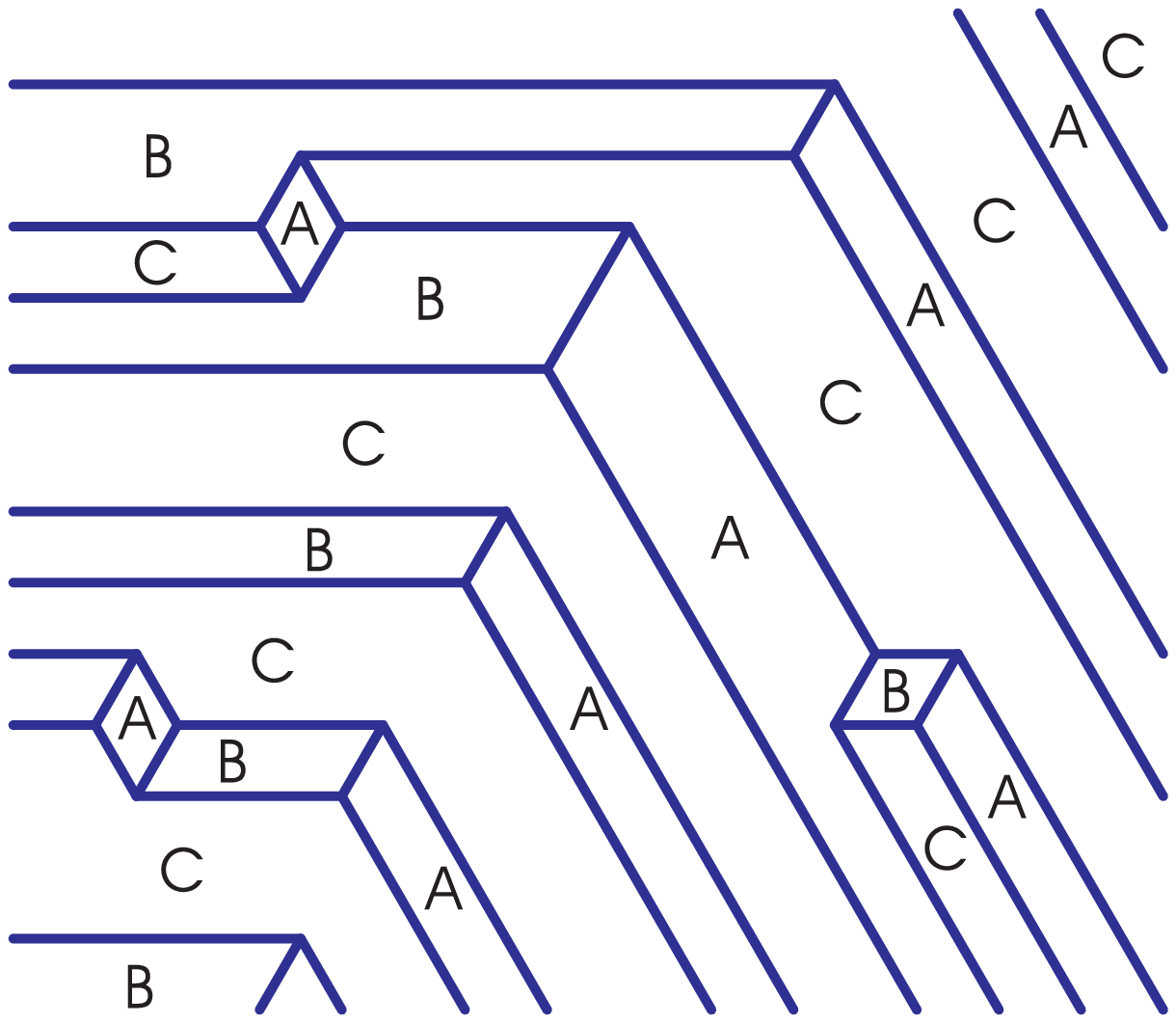}
\caption{A boundary between two uniaxial networks with different
preferable orientations of domain walls.}

\end{figure}

A boundary between two uniaxial-network states with different preferable
orientations of domain walls is schematically shown in Fig.
\ref{Boundary}, where each line corresponds to a domain wall, whereas
letters A, B and C again denote domains with three different orientation
of stripes. Note that one of the uniaxial networks (the one to the right)
contains mainly domains A and C, whereas the other one domains B and C. It
is clear that the change of the dominant orientation of domain walls which
has to take place at a boundary between domains requires to have on this
boundary a large concentration of fractional vortices and accordingly the
energy of such a boundary per unit length has to be of the order of $E_0$.

Accordingly, the expression for the free energy of the domain wall network
formed by unpaired fractional vortices separated by distances of the order
of $L$, must also contain a positive contribution of the order of
${E_{0}}/{L}$, which for large $L$ will always dominate over the first
term in Eq. (\ref{F_NW}). This suggests that the phase transition related
to the appearance of free fractional vortices must take place at $T=T_{\rm
dis}>T_{\rm FV}$ and has to be of the first order, because the two minima
of the function $F_{\rm NW}(L)$ (at $L=\infty$ and at finite $L$) will
always be separated by some barrier. Moreover, since the behavior of the
function $F_{\rm NW}(L)$ is determined (in addition to $T$) by the single
energy scale, $E_0$, one can expect that the shift of the transition
temperature, $T_{\rm dis}$, with respect to $T_{\rm FV}$ will be of the
order of $E_0$, whereas the typical distance between fractional vortices
just above the transition can be only numerically larger than one, and not
parametrically. Accordingly, we can conclude that $T_{\rm dis}\sim
(0.05\div 0.1) J$, but cannot provide a more accurate estimate.

Note that the formation of a domain-wall network leading to the
intermixing of all six striped vortex patterns is impossible without the
proliferation of free fractional vortices. This means that there is no
possibility for the phase transitions related with the removal of the
triple degeneracy of the uniaxial network states and suppression of the
algebraic phase behavior of $C_8({\bf r})$ to take place at different
temperatures. On the other hand, the appearance of free fractional
vortices leads to the screening of the logarithmic interaction of the
conventional vortices, so that they also unbind. Accordingly, at $T>T_{\rm
dis}$, the system is in the completely disordered phase in which all
correlations decay exponentially.

\section{Conclusion\label{conc}}
\subsection{Phase diagram}

The main results of this work are summarized in Table I reviewing  the
properties of various phases represented in the phase diagram of the
fully-frustrated XY model on a honeycomb lattice.  The order in which
these phases are listed corresponds to the increase in temperature. By the
existence of the long-range order (LRO) in terms of vortex pattern we mean
the selection of one of the six equivalent striped vortex patterns, see
Fig. \ref{GroundStates}(a). On the other hand, the existence of the
long-range order in terms of the domain-walls orientation implies the
choice of one of the three preferable orientations of domain walls. The
last three columns describe the behavior of the gauge-invariant phase
correlation functions defined by Eq. (\ref{C_p}).

The first line in Table I refers to the behavior of the correlation
functions when the averages are calculated over the set of the all ground
states of the model. A typical ground state incorporates an irregular
sequence of parallel straight domain walls, which leads to the exponential
decay of the chirality correlation function, Eq. (\ref{Cch}), as well as
of phase correlation functions $C_p({\bf r})$ with $p<8$ for all
directions that are not exactly perpendicular to one of the directions of
lattice bonds. Nonetheless, the system has the long-range order in terms
of domain-walls orientation

At $T>0$ domain walls acquire a finite free energy per unit length given
by Eq. (\ref{fDW}) and originating from the difference in the free energy
of spin waves. \cite{KD} At the lowest temperatures this leads to the
suppression of infinite domain walls crossing the whole system and the
appearance of the long-range order in term of vortex pattern (phase No. 1
in Table I). The system spontaneously chooses one of the six equivalent
striped vortex patterns. On the other hand, all phase correlation
functions decay algebraically due to the presence of spin waves.

With the increase in temperature, the first phase transition takes place
at \makebox{$T=T_{\rm DDDW}\approx 0.8\times10^{-2}\,J$} and is related to
the appearance of a sequence of fluctuating double-spaced double domain
walls. 
On the average, the direction of these fluctuating walls is perpendicular
to the direction of vortex stripes. This phase transition is continuous
with the Ising-type behavior in a very narrow critical region and the
Pokrovsky-Talapov behavior in a wider region around the critical one.
Although the transition taking place at $T=T_{\rm DDDW}$ is related to the
appearance of domain walls, above it (phase No. 2 in Table I) the discrete
degeneracy (related to the long-range order in terms of vortex pattern)
remains the same (six-fold), whereas the decay of the phase correlation
function $C_1({\bf r})$ changes from algebraic to exponential. This
happens because the presence of a doubly-spaced double domain wall between
two points shifts the phase difference by $\pi$.

The next phase which appears with the increase in temperature (phase No. 3
in Table I) has the same symmetry properties as a typical ground state of
the model. In particular, in this phase there is no long-range order in
terms of vortex pattern. The main difference with a random sequence of
parallel straight domain walls (characteristic for a typical ground state)
consists in the presence of local defects which shift the positions of
these walls but do not change their preferable orientation, see Fig.
\ref{Network}(b). The phase transition to this uniaxial domain-wall
network  has to be of the first order and can be expected to take place
when $T$ is around $2T_{\rm DDDW}$. The important feature of the uniaxial
domain-wall network is that the nodes of this network, which in terms of
phase variables correspond to logarithmically interacting fractional
vortices, are bound in neutral pairs.

The last phase transition leading to the complete suppression of the
long-range order and to exponential decay of all correlation functions can
be associated with the appearance of an isotropic domain-wall network in
which some fractional vortices (the nodes of the network) are not bound in
pairs but free. By analyzing how the free energy of such a network depends
on typical distance between free fractional vortices it turns out possible
to establish that this transition must be of the first order and has to
occur at $T_{\rm dis}\sim (0.05\div 0.1) J$.

All phase transitions mentioned above take place at temperatures which are
much lower than the temperature $T_{\rm BKT}\sim J$ at which the unbinding
of pairs of the conventional vortices (with integer topological charges)
would take place if any other fluctuations were absent. The existence at
such temperatures of an exponentially small concentration of small bound
pairs of integer vortices does not lead to any noticeable consequences. On
the other hand, at $T>T_{\rm dis}$ the logarithmic interaction of integer
vortices is screened due to the presence of a finite concentration of free
fractional vortices and therefore there are no grounds to expect the
existence of an additional phase transition related to the unbinding of
integer vortices.

\subsection{Discussion}

Thus we have established that the equilibrium thermodynamics of the fully
frustrated XY model on a honeycomb lattice supports the existence at $T>0$
of four different phases three of which are characterized by the presence
of a long-range order. However the observation of some of these phases
(and phase transition between them) in a real or numerical experiment may
turn out to be a big problem.

In particular, at $0<T<T_{\rm DDDW}$, the complete suppression of domain
walls crossing the whole system (induced by the spin-wave contribution to
their free energy, see Sec. \ref{SpinWaves})  is achieved only in the
thermodynamic limit, $L_{\rm max}\to \infty$. In a finite system with
linear size $L_{\rm max}$, in order to have the average number of such
walls smaller than one, the size of the system should be larger than
$L_c(T)$, the temperature-dependent solution of the equation
\begin{equation}                                               \label{Lc}
    L_{\rm max}\exp\left(-\gamma\frac{TL_{\rm max}}{J}\right)=1
\end{equation}
At the point of the phase transition related to the appearance of infinite
double-spaced double domain walls, \makebox{$L_c\approx 3\times 10^7$}.
This means that numerical simulations of systems with $L\lesssim 10^7$
does not allow one to observe the phase with suppressed domain walls
independently of whether it is possible or not to reach thermal
equilibrium at $T\sim T_{\rm DDDW}$.

\begin{widetext}

\begin{table}[bp]
\begin{tabular}{|c|c|c|c|c|c|c|}
\hline 
  \multicolumn{2}{|c|}{} & \multicolumn{2}{|c|}{LRO in terms of}
& \multicolumn{3}{|c|}{decay of phase} \\
\multicolumn{2}{|c|}{}  & vortex & domain walls &
\multicolumn{3}{|c|}{correlations}   \\
\cline{5-7} \multicolumn{2}{|c|}{}  & pattern & orientation &
 $~C_1({\bf r})~$ & $~C_2({\bf r})~$ & $~C_8({\bf r})~$ \\
\hline \hline $T=0$ & straight parallel domain walls
& $-$ & $+$ & exp & exp & LRO \\
\hline 1 & no infinite domain walls& $+$ & $-$ & a & a & a \\
\hline 2 & a sequence of fluctuating
DDDWs & $+$ & $-$  & exp & a & a \\
\hline 3& uniaxial domain-wall network & $-$ & $+$  & exp & exp & a
\\
\hline 4 & disordered phase & $-$ & $-$  & exp & exp &
exp\\
\hline

\end{tabular}
\vspace*{6mm} \caption{The properties of different phases of the fully
frustrated XY model on a honeycomb lattice. In the last three columns
``exp" denotes an exponential decay of the corresponding correlation
function (for a generic direction) and ``a" an algebraic decay. }
\end{table}
\newpage
\end{widetext}

More or less the same criterion applies to the observation of the phase
with a sequence of fluctuating double domain walls. Accordingly, for
$L\lesssim 10^7$ the increase of temperature will lead just to  a smooth
crossover from a

\hspace*{-3.5mm}random sequence of parallel straight domain walls to the
uniaxial domain-wall network with the same preferable orientations of
domain walls. However, a discontinuous transition from the uniaxial
domain-wall network to an isotropic one (the disordered phase) can be
expected to be observable even if the size of the system is not so large.
On the other hand, it follows from the numerical simulations of Ref.
\onlinecite{RBM} that the task of reaching thermal equilibrium at
$T\lesssim 0.1\, J$ may require special efforts for its solution.

In the case of the Berezinskii-Villain interaction the free-energy of spin
waves is exactly the same for all domain-wall configurations and,
therefore, does not lead to the removal of the accidental degeneracy of
the ground states. In such a situation, the scenario of a smooth crossover
from a random sequence of parallel straight domain walls to the uniaxial
domain-wall network works not only at $L_{\rm max}\lesssim L_c$ bit also
in the thermodynamic limit. That is, instead of three phase transitions an
infinite system with the Berezinskii-Villain interaction must experience
only one, related to the loss of the long-range order in the orientation
of domain walls and of the algebraic decay of $C_8({\bf r})$. The
conclusion that this transition has to be of the first order is not
sensitive to a particular form of the interaction in the Hamiltonian of
the model.

Since the fully-frustrated XY model with the Berezinskii-Villain
interaction on a honeycomb lattice is exactly equivalent to the
half-integer Coulomb gas on a triangular lattice, the scenario described
in the previous paragraph has to be realized also in such a Coulomb gas.
Numerical simulations of the half-integer Coulomb gas on a triangular
lattice were undertaken by Lee and Teitel, \cite{LT} who have discovered a
relatively sharp jump of the dielectric function at $T/J\sim 0.04$
but assumed it to be a finite-size induced artefact. This conclusion was
based on the observation that the formation and motion of finite-energy
local defects (analogous to those shown in our Fig. \ref{Rhombi})
destroys the long-range order in vortex pattern at an arbitrary low
temperature. However, this argument missed a possibility of having the
long-range order in domain-walls orientation, which according to our
analysis should be present in the half-integer Coulomb gas on a triangular
lattice at low enough temperatures.

It should be mentioned that a uniformly frustrated $XY$ model with this or
that interaction provides just an idealized description of a
Josephson-junction array or of a superconducting wire network. In physical
situations there can exist other mechanisms for the removal of the
accidental degeneracy of the ground states related to the interactions not
taken into account in the framework of an XY model. One of them is the
magnetic interaction of currents in the junctions or in the wires.
However, it is likely that the sign of the domain-wall energy induced by
this interaction may depend on the particular geometry of the system.

If $E_{\rm DW}$, the domain wall energy per unit length, is positive, this
will substantially improve the possibilities for the observation of the
phase with the striped vortex pattern. Even if $E_{\rm DW}$ is few orders
of magnitude smaller than $J$, it will be many orders of magnitude larger
than the value of $f_{\rm DW}$ given by Eq. (\ref{fDW}) at the
corresponding temperatures. The opposite sign of $E_{\rm DW}$ will lead to
the stabilization of another periodic vortex pattern shown in Fig. 1(c) of
Ref. \onlinecite{KD}. In any case, a nonzero value of $E_{\rm DW}$ will
allow the observation of a phase with a long-range order in terms of
vortex  pattern in the systems of less than macroscopic sizes.

On the other hand, the magnetic interaction of currents leads to the
screening of the logarithmic interaction of vortices (both conventional
and fractional) at large distances. This will transform the phase
transition related to the appearance of a sequence of fluctuating
double-spaced double domain walls into a crossover, because when the
logarithmic interaction of fractional vortices is screened, such double
walls no longer are topologically stable defects and can have free
end-points. A substantial increase of $E_{\rm DW}$ may lead also to the
disappearance of the region of stability of the uniaxial network state,
which implies a direct transition from the phase with the long-range order
in terms of vortex pattern into the disordered phase.

Another mechanism for the stabilization of striped vortex patterns in
magnetically frustrated superconducting wire networks may be related with
the non-uniformity of the order parameter amplitude \cite{Xiao} (see also
Ref. \onlinecite{Sato}). However the conclusions of Ref. \onlinecite{Xiao}
are in some contradiction with the results of Ref. \onlinecite{EZh}
devoted to including into analysis the higher-order terms of the
Ginzburg-Landau expansion, which suggests that the problem requires
further investigation.

The uniformly frustrated XY model with triangular lattice and
$f=\frac{1}{4}$ also allows for the formation of zero-energy domain walls
parallel to each other, \cite{KVB} so its phase diagram may have some
common features with the one constructed in this work. In the uniformly
frustrated XY model with $f=\frac{1}{3}$ on a triangular lattice
\cite{KVB}, as well as in the fully frustrated XY model on a dice lattice
\cite{FF-dice-01,FF-dice-05} the situation is more complex because these
models allow for the formation of {two} sets of parallel zero-energy
domain walls which can cross each other without paying any energy for
this. However, in such a situation it is also possible to expect the
existence of a phase in which vortex pattern is disordered whereas the
long-range order is related to the orientation of domain walls.

On the experimental side, the conclusions of this work may be applicable
not only to Josephson-junction arrays and superconducting wire networks
with a half-integer number of flux quanta per plaquette, but also to
magnetically frustrated triangular arrays of microholes \cite{Yoshida} or
nanoholes \cite{Stewart} in thin superconducting films. For a half-integer
number of flux quanta per hole, the energy of different vortex
configurations in such objects has to be described by the Hamiltonian of
the half-integer Coulomb gas with screened logarithmic interaction.

\acknowledgments

This work was supported in part by the Russian Fund for Basic Research
Grant No. 09-02-01192a.
\appendix

\section{Berezinskii-Villain interaction and the Coulomb gas representation
\label{BV}}

A magnetically frustrated network formed by identical superconducting
wires can be described in London regime (when the amplitude of the
superconducting order parameter is uniform along the wires) by the
Hamiltonian \cite{AII}
\begin{subequations}                                         \label{Hnw}
\begin{equation}                                              \label{HBV}
    H=\frac{J}{2}\sum_{\left(jj'\right)}\theta^2_{jj'}
\end{equation}
where $\theta_{jj'}$ is the integral of the gauge-invariant phase gradient
along the wire connecting nodes $j$ and $j'$. Variables $\theta_{jj'}$ can
acquire arbitrary values, $-\infty<\theta_{jj'}<+\infty$, but on any
plaquette of the lattice have to satisfy the constraint
\begin{equation}                                              \label{f-2}
    \sum_{\Box\,\alpha} \theta_{jj'}=-2\pi f\;\;(\mbox{mod }2\pi)\;,
\end{equation}
\end{subequations}
where parameter $f$ depends on the applied magnetic field and is equal to
the number of superconducting flux quanta per plaquette.

In terms of phase variables $\varphi_j$ defined on the nodes of the
network, the partition function of the model (\ref{Hnw}) can be rewritten
\cite{VKB} as the partition function of the uniformly frustrated XY model
with the so-called Berezinskii-Villain interaction $V_{\rm BV}(\theta)$
defined by the relation
\begin{equation}                                     \label{V_BV}
\exp\left[-\frac{V_{\rm BV}(\theta)}{T}\right]
=\sum_{h=-\infty}^{\infty}\exp \left[-\frac{J}{2T}(\theta-2\pi
h)^2\right]\;.
\end{equation}
The function $V_{\rm BV}(\theta)$ has the same symmetry and periodicity as
$V(\theta)=-J\cos\theta$. For $J\gg T$ it is everywhere except the close
vicinity of the point $\theta=\pi$ close to parabola,
\makebox{$V_{\rm BV}(\theta)\approx({J}/{2})\theta^2+\mbox{const}$}.
In the opposite limit, $J\ll T$, the function $V_{\rm BV}(\theta)$ with an
exponential accuracy is reduced to \makebox{$-J_{\rm
eff}\cos\theta+\mbox{const}$}, where however coupling constant
$J_{\rm eff}=2T\exp(-T/2J)$
is much smaller than $J$.

The interaction function defined by Eq. (\ref{V_BV}) was introduced by
Berezinskii \cite{Ber} and Villain \cite{Vil75} because it allows to simplify the
analytical analysis of XY models. Namely,
the integration over variables $\varphi_j$ in the partition function of an
XY model with such an interaction is Gaussian and therefore can be
performed exactly. In particular, in the case of the periodic boundary
conditions the application of this procedure transforms the partition
function of a uniformly frustrated XY model with the Berezinskii-Villain
interaction into that of a ``fractional'' Coulomb gas \cite{FHS} described
by the Hamiltonian
\begin{equation}                                               \label{HCG}
H_{\rm CG}=\frac{1}{2}\sum_{\alpha,\beta}
m_{\alpha}G_{\alpha\beta}m_{\beta}\;,
\end{equation}
where variables $m_{\alpha}$ (the charges of the Coulomb gas) are defined
on the sites $\alpha$ of the dual lattice and acquire values shifted with
respect to integers by $-f$. Each of these variables is proportional to
the sum of variables $\theta_{jj'}-h_{jj'}$ (defined on lattice bonds)
over the perimeter of plaquette $\alpha$ and, in terms of the XY
representation, can be identified with the vorticity of this plaquette
divided by $2\pi$.

The long-range interaction $G_{\alpha\beta}$ entering Eq. (\ref{HCG}) has
a form
\begin{equation}                                     \label{G0}
G_{\alpha\beta}
=4\pi^2 J (-\hat{\Delta})^{-1}_{\alpha\beta}\;,
\end{equation}
where $\hat{\Delta}$ is the discrete Laplace operator on the dual lattice.
This interaction logarithmically depends on \makebox{$r_{\alpha\beta}=
|{\bf r}_\alpha-{\bf r}_\beta|$,} the distance between $\alpha$ and
$\beta$,
\begin{equation}                                     \label{G0R}
G_{\alpha\alpha}-G_{\alpha\beta}= 2\pi \Gamma_0\left(\ln{
{r}_{\alpha\beta}}
+\kappa\right)\;, \vspace*{1mm}
\end{equation}
where $\Gamma_0$ is proportional to $J$ but also depends on the structure
of the lattice. In terms of the XY representation $\Gamma_0$ is the
helicity modulus describing the rigidity of the system with respect to
phase twist. In the considered case 
$\Gamma_0=J/\sqrt{3}$, whereas $\kappa=\pi/\sqrt{3}$ for $r_{\rm
{\alpha\beta}}=1$ (the nearest neighbors on the triangular lattice) and
monotonically increases by less than 0.5\% with the increase of
${r}_{\alpha\beta}$ to infinity.

It can be expected that the number of qualitative features of frustrated
XY models with the conventional and with the Berezinskii-Villain
interactions are the same. However, in situations when the ground states of
the model possess an accidental degeneracy, one should be cautious because
in a model with the Berezinskii-Villain interaction the free energy of the
spin waves is exactly the same for all vortex configurations and therefore
does not lead to the removal of such a degeneracy (in contrast to the
models with the conventional form of the interaction). 

In the case of a fully-frustrated model (with $f=\frac{1}{2}$), the
charges of the Coulomb gas are half-integer. Since the charges of the
opposite signs attract each other it is rather clear that when the dual
lattice is bipartite (square or honeycomb) the minimum of energy is
achieved in configurations with a regular (checkerboard-like) alternation
of positive and negative charges. This is in a perfect agreement with
having a two-fold discrete degeneracy of the ground states in terms of the
XY representation. In the case of the half-integer Coulomb gas on a
trangular lattice it is impossible to construct a configuration in which
each charge has neighbors only of the opposite sign. \makebox{Accordingly,
it is much less evident what are the states}

\begin{widetext}
\hspace*{-3.5mm}minimizing the energy and how high is their degeneracy. In
Appendix \ref{CG} we propose a transformation which substantially
simplifies this task and also provides a transparent expression for the
energies of the excited states of the half-integer Coulomb gas.

\section{Effective charge representation for the energy of the
half-integer Coulomb gas \label{CG}}

When considering the half-integer Coulomb gas described by Hamiltonian
(\ref{HCG}) it is convenient to introduce auxiliary variables $q_\alpha$
linearly related to $m_\alpha$,
\begin{equation}                                             \label{q}
    {q_\alpha}=m_\alpha+\lambda\hat{\Delta}_{\alpha\beta}m_\beta\;,
\end{equation}
where $\hat{\Delta}_{\alpha\beta}$ is the discrete Laplacian on the
lattice on which variables $m_\alpha$ are defined and the choice of
$\lambda$ depends on the structure of this lattice. The replacement of
$m_\alpha$ by $q_\alpha-\lambda\hat{\Delta}_{\alpha\beta}m_\beta$ allows
to rewrite Eq. (\ref{HCG}) as
\begin{equation}                                             \label{HCG-2}
    H_{\rm CG}=
    \frac{1}{2}\sum_{\alpha,\beta}q_\alpha G_{\alpha\beta}q_\beta
    +4\pi^2 J\left(\lambda-{\lambda^2 z}\right)\sum_\alpha m_\alpha^2
    +2\pi^2 J\lambda^2\sum_{(\alpha\alpha')}(m_\alpha+ m_{\alpha'})^2
    \;,
\end{equation}
\end{widetext}
where $z$ is the coordination number of the lattice, whereas the summation
in the last term is performed over all pairs of nearest neighbors
$(\alpha\alpha')$. The form of Eq. (\ref{HCG-2}) suggests that the
long-range interaction of the charges of the half-integer Coulomb gas can
be replaced by their repulsion on neighboring sites plus the long-range
interaction of the ``effective charges" $q_\alpha$ defined by Eq.
(\ref{q}). Naturally, it is convenient to choose the value of $\lambda$ in
such a way that in the ground states all effective charges are equal to
zero, which substantially simplifies the calculation of energy and also
minimizes the first term in Eq. (\ref{HCG-2}).

For $\lambda<{1}/{z}$ the second term in Eq. (\ref{HCG-2}) is minimized
when $m_\alpha=\pm \frac{1}{2}$, whereas the third term requires to
maximize the number of bonds on which $m_\alpha$ and $m_{\alpha'}$ have
opposite signs. On bipartite lattices it is always possible to do this for
any pair of neighboring sites. In such a case, the fulfillment of
condition $q_\alpha=0$ is achieved when one takes $\lambda={1}/{(2z)}.$
For square lattice this gives $\lambda=\frac{1}{8}$ and for honeycomb
lattice $\lambda=\frac{1}{6}$. For these values of $\lambda$ the states
with the regular checkerboard alternation of positive and negative charges
$m_\alpha=\pm 1$ ensure the simultaneous minimization of all three terms
in Eq. (\ref{HCG-2}) which rigorously proves that they are the ground
states.

In the case of a triangular lattice, each plaquette has to have either one
or three bonds with the same sign of $m_\alpha$ and $m_{\alpha'}$. As a
consequence of this, on the average, each charge has to have at least two
neighbors of the same sign. In such a situation it is convenient to take
$\lambda=\frac{1}{8}$, which for all configurations  with $m_\alpha=\pm
\frac{1}{2}$  reduces Eq. (\ref{HCG-2}) to
\begin{equation}                                             \label{HCG-3}
    H_{\rm CG}=
    \frac{1}{2}\sum_{\alpha,\beta}q_\alpha G_{\alpha\beta}q_\beta
    +\left(N+\frac{N_{3}}{2}\right)g
    \;,
\end{equation}
where
\begin{equation}                                               \label{g}
g = \left(\frac{\pi}{4}\right)^2J\;,
\end{equation}
$N$ is the total number of sites and $N_{3}$ is the number of triangular
plaquettes on which all three charges have the same signs. It is evident
that the first term in Eq. (\ref{HCG-3}) is minimized when all variables
$q_a$ are equal to zero and the second one when $N_{3}=0$. Since both
these conditions can be fulfilled simultaneously, constant $g$ is nothing
else but the ground-state energy per site. Note that the value of $g$
given by Eq. (\ref{g}) is in perfect agreement with having in the fully
frustrated XY model $\theta_{jj'}=0$ on one third of the bonds of the
honeycomb lattice and \makebox{$\theta_{jj'}=\pm \pi/4$} on all other
bonds.

Thus, the set of the ground states of the half-integer Coulomb gas
includes all states in which all charges are equal to $\pm 1/2$, each
charge has exactly two neighbors of the same sign and on each triangular
plaquette there are always two charges of one sign and one charge of the
opposite sign. These states can be put into the correspondence with the
ground states of the fully frustrated XY model on a honeycomb lattice by
interpreting the charges $m_\alpha$ as vorticities (divided by $2\pi$) of
the corresponding plaquettes, as it could be expected from the equivalence
between the half-integer Coulomb gas and the fully frustrated XY model
with the Berezinskii-Villain interaction on the dual lattice
(see Appendix \ref{BV}). However, the reduction of Hamiltonian (\ref{HCG})
to the form (\ref{HCG-3}) has allowed us to find the structure of the
ground states of the half-integer Coulomb gas on a triangular lattice
directly in terms of the Coulomb gas representation.

Note that the form of Eqs. (\ref{HCG-2}) and (\ref{HCG-3}) relies on the
very special form of the interaction of charges in the Coulomb gas [given
by Eq. (\ref{G0})]. As soon as the form of the interaction is modified
(for example, to take into account the screening effects in a
superconducting array or wire network), the possibility to transform the
Hamiltonian to the form (\ref{HCG-2}) with only local interaction of
variables $m_\alpha$ disappears, which leads to a partial removal of the
degeneracy between the states with $q_\alpha=0$ and $N_{3}=0$.

In the configurations with $m_\alpha=\pm \frac{1}{2}$ on a triangular
lattice, the values of effective charges $q_\alpha$ are given by
\begin{equation}                                                 \label{q-2}
    q_\alpha=\frac{N_\alpha-2}{4}m_\alpha\;,
\end{equation}
where $N_\alpha$ is the number of the nearest neighbors of site $\alpha$
with the same sign of the charge $m$. It follows from Eq. (\ref{q-2}) that
the effective charges of various sites are either zero or equal to
$\pm\frac{1}{8}$ or multiples of these values. In Sec. \ref{FV} the
appearance of logarithmically interacting vortices with fractional
topological charges on those plaquettes of the honeycomb lattice which
have the number of neighbors with the same sign of vorticity not equal to
two has been discussed directly in terms of the XY model.

Eq. (\ref{HCG-3}) provides a convenient way for calculating the energies
of local defects in half-integer Coulomb gas on a triangular lattice (or
in the equivalent fully frustrated XY model with the Berezinskii-Villain
interaction on a honeycomb lattice) by summing just few terms. The values
of these energies for various finite-energy defects discussed above and
illustrated in Figs. \ref{Pairs-a}-
\ref{Intersections} are listed in Table II.

\begin{table}[b]
\begin{centering}
\begin{tabular}{|c|c|c|}
\hline figure & defect energy & notation\\
\hline 
\ref{Pairs-a}(a) & 0.102808 & \\
\ref{Pairs-a}(b) & 0.134471 & \\
$~$\ref{Pairs-b}(a), \ref{Pairs-b}(b), \ref{Pairs-b}(c)$~$ & 0.142292 & $E_{\rm B}$\\
\ref{Intersections}(b) & 0.197419 & \\
\ref{Intersections}(a) & 0.237279 & \\
\ref{Rhombi}(a), \ref{Rhombi}(b)  & 0.253298 & $E_{\rm R}$\\

\ref{EndPoints}(b) & 0.347909 & \\
\ref{Kinks}(a) & 0.371750 & $E_{\rm K}$ \\
\ref{Kinks}(c) & 1.977200 & $E_{\rm K}^{\rm s}$\\
\hline

\end{tabular}
\caption{The energies (in units of $J$) of the finite-energy defects
presented in Figs.
\ref{Pairs-a}-
\ref{Intersections} for the case of the Berezinskii-Villain interaction.}
\end{centering}

\end{table}


\begin{thebibliography}{99}
\bibitem{comm-anisotropy} When the helicity modulus is an anisotropic
               tensor, $\Gamma$ in Eq. (\ref{E_FV}) stands for the
               geometric mean of its eigenvalues.
\bibitem{renorm} This estimate does not take into account the
                 renormalization effects (related to the presence of bound
                 pairs of fractional vortices), which would shift
                 $T_{\rm FV}$ further down.
\end{thebibliography}
\end{document}